\documentclass[useAMS,usenatbib,onecolumn]{mn2e}
\usepackage{natbib,aas_macros}

\newcommand{\dD}[2]{\frac{\mathrm{d}{#1}}{\mathrm{d}{#2}}} 
 
\usepackage{amsmath,amssymb}
\usepackage[dvipdfm]{graphicx,color}
\title[A Particle Simulation for the Pulsar Magnetosphere: II]
 {A Particle Simulation for the Axisymmetric Pulsar Magnetosphere:
II. the case of dipole field}
\author[T. Wada and S. Shibata]{%
Tomohide Wada$^1$\thanks{E-mail: tomohide.wada@nao.ac.jp}
\& Shinpei Shibata$^2$\thanks{E-mail: shibata@sci.kj.yamagata-u.ac.jp}%
\\
$^1$National Astronomical Observatory of Japan,
Osawa 2-21-1, Mitaka, 181-8588, Japan\\
$^2$Department of Physics, Yamagata University,
Kojirakawa, Yamagata 990-8560, Japan}
\begin{document}

\date{Accepted 2011 July 26. Received 2011 July 25; in original form 2010 November 22}

\pagerange{\pageref{firstpage}--\pageref{lastpage}} \pubyear{2009}

\maketitle

\label{firstpage}

\begin{abstract}
The main issue of the pulsar magnetosphere is how the
rotation power is converted into both particle beams which
causes pulsed emissions, and a highly relativistic
wind of electron-positron plasmas which forms surrounding
nebulae shining in X-rays and TeV gamma-rays.
As a sequel of the first paper (\citealt{2007MNRAS.376.1460W}), 
we carried out a three dimensional particle simulation 
for the axisymmetric global magnetosphere.
We present the results of additional calculations, which are higher
 resolution model and higher pair creation rate cases, and a detailed
 analysis for the solution.
We confined to demonstrate the cases of low pair creation rate, i.e.,
 the magnetic field is fixed dipole. 
The radiation drag of the plasma is taken in a form with the
 curvature radius along the dipole magnetic field.
The electrostatic interactions are calculated by a programmable
special purpose computer, GRAPE-DR \citep{Makino2007}.
Once pair creation is onset in the outer gaps, the both signed particles
 begin to drift across the closed magnetic field due to radiation drag,
 and they create outflow.
Eventually, the steady magnetosphere has outer gaps,
 both signed outflow of plasma and a region in which the electric
 field is dominant extending from the equator.
In the steady state, the magnetic field made by magnetospheric current
 is comparable to the dipole magnetic field outside of several light
 radii from the star.
In much more pair creation rate model, the effect of modification of the
 magnetic field will bring about modification
 of the outflow of the plasma, requiring further study with
 higher pair creation rate model in a subsequent paper.
\end{abstract}


\section{Introduction}
\label{chap:9ruztWG4}

Rotation powered pulsars are bright sources in the X-ray sky
(e.g., \citealt{1997A&A...326..682B,2002A&A...387..993P,2008AIPC..983..171K})
 and the gamma-ray sky 
(e.g., \citealt{1996ApJ...465..898N,2008arXiv0803.0116D}) as well
as in the radio
(e.g., \citealt{2004MNRAS.352.1439H}), and they are identified as
 magnetized rotating neutron stars. 
The pulsed emissions originate in the accelerated particles in the
magnetosphere.
In addition, as a persistent source of high energy photons, we observe
nebulae around young pulsars with synchrotron X-ray radiation
\citep{2008AIPC..983..171K}, inverse Compton TeV radiation
\citep{2008arXiv0803.0116D}, and thermal radiations from the neutron
star \citep{1997A&A...326..682B}.
The emission from the nebula is caused by termination shock of the
pulsar wind, which is believed to be relativistic outflow of 
magnetized pair plasmas carrying out most of the rotation power
\citep{1974MNRAS.167....1R,1984ApJ...283..694K,1984ApJ...283..710K}.
However, the nature of the pulsar wind and the radiation beam from the
pulsar magnetosphere have remained a longstanding problem.
As a primitive model, \citealt[hereafter GJ]{1969ApJ...157..869G},
suggested an axisymmetric steady model of pulsar magnetosphere in which
the neutron star is assumed to be a perfect conductor surrounded
everywhere by charge-separated force-free plasmas.
The model anticipates that the plasma holds corotation with the star in the
light cylinder although this corotational magnetosphere can be violated
beyond the light cylinder.
Then the GJ model assumes that the plasma is replaced by the acceleration
along the opened magnetic field line beyond the light cylinder.
On the basis of the copious plasma in the magnetosphere,
the global relativistic magnetodynamics models are applied to help us
understand acceleration of the pulsar wind 
\citep{2006MNRAS.367.1797M,2006MNRAS.368L..30M,2006MNRAS.367...19K,2006MNRAS.368.1717B}
and possible dissipation such as in magnetic neutral sheets may 
contribute to high energy pulsed radiation
\citep{2002A&A...388L..29K}.
Against these studies, several studies showed that an aligned rotating
neutron star should consist of the electrostatic charge clouds
surrounded by the vacuum gaps without pulsar wind and therefore the
steady solutions with outflow of the plasma are not self-consistent 
(See, \citealt{2001RMxAC..10..168M}).
About the electrostatic solution, 
\citet[hereafter KM]{1985MNRAS.213P..43K,1985A&A...144...72K}
performed a particle simulation for the axisymmetric model and showed
formation of the gaps around the null surface with a rotating equatorial
disc and polar domes. 
This model suggests a quiet electrosphere when the plasma is extracted
only from the stellar surface, but the vacuum gap in the middle
latitudes looked to be unstable against the pair creation cascade. 
Successively, \cite*{2001MNRAS.322..209S} reconstructed the same result
with a higher accuracy and even if the pair plasma were generated in the
gaps at the middle latitudes, the quiet solution without pulsar wind was
identified.
In their models, the quiet solution is obtained that the particles are
put at equilibrium positions along the dipole magnetic field line and
this picture would be valid if the plasma is set down fully inside the
light cylinder and the magnetic field is larger than the electric field.
In other words, the effects of the inertia of the co-rotational plasma
around the light cylinder is not considered.
The same result is obtained by \cite*{2002A&A...384..414P} who concluded
a given total charge of the system determined by the net outgoing
charged flux and the potential configuration is unfavorable for the
particles to escape to infinity such as the active solution obtained by
MHD approximation.
However, the quiet solution should be modified to concern the leaking of
the plasma at which the electrostatic plasma is in the
vicinity of the light cylinder.
As a possibility of radial outflow from the equatorial disc on the
equator, \cite{2004IAUS..218..357S} discovered the leaking mechanism
of the equatorial disc caused by the diocotron instability with numerical
simulation, which is also pointed out by \cite{2002A&A...387..520P} with
linear theory.
The leaking of the disc should be significant to make the current in the
vacuum region outside of the quiet electrosphere.
But the time scale of the diffusion of the plasma in the vicinity of the
light cylinder is much larger than the rotation period and
therefore further study is needed to follow the global current structure
with a much longer time scale such as a global numerical simulation with
much more computational time with relativistic kinematics.

Contrary to the previous quiet solutions, we reported an active
solution  
\citep[hereafter, WS]{2007MNRAS.376.1460W} with coexisting pulsar wind
and outer gaps in which the effect of the inertia and radiation drag for
the plasma is included by 3-dimensional particle simulation with special
relativistic regime, and the simulation starts from the quiet solution
as KM.
Once pair plasmas are generated in the vacuum gaps around the quiet
electrosphere, the positive charged outflow is formed near the equator
and the negative charged outflow is formed from the polar region of the
star.
The outflow of plasma maintains the charge deficiency in the outer gaps.
For the balancing of loss and the supply of particles from the outer
gaps, the magnetosphere eventually settles in a steady state with pulsar
wind.
But there remains some problems to solve.
At first, although our simulation showed that the wind coexists with the
pair creating regions in the pulsar outer magnetosphere, the solution is
only obtained by a lowest case of pair creation rate, where we
carried out the simulation in which the current density around the star
is less than the one of GJ model, and therefore magnetic field by the
magneto-spheric current is negligibly-small in the light cylinder. 
If the pair creation rate taken to be large gradually in our simulation,
the gaps tend to reduce and decrease the generation rate of the plasma
and therefore the active structure might be disappeared.
Although the drift motion of plasma caused by radiation
drag force works as a leaking mechanism in closed magnetic field and is
favorable to maintain the outer gaps, recalculation should be needed
whether our active solution is maintained in higher pair creation rate
models.
Secondly, the obtained structure of the outer gaps contains some
artifact due to the accuracy.
The detailed flow pattern of plasma such as the poloidal current loop
inside of the light cylinder was not clear because of low resolution of
our previous work. 
We will simulate the lowest pair creation rate model again with higher
resolution in which the unit charge of the particle is taken to be
$1/10$.
At last, our particle simulation has numerical error, which is defined
as the resolution of charge of simulation particle and the injection of
the plasma from the inner boundary.
In our previous work (WS), the polar potential drop is maintained in
three times larger fraction than artifact level, present higher
resolution model will decrease the error and could identify the existence.
However the particle has large inertia length
which is about $3\%$ of the stellar radius and therefore we denote that
our simulation could not resolve the size of polar cap accelerator. 

Meanwhile the development of global magnetosphere, some phenomenological
models which concerns the local structure for the magnetospheric
activities have been developed:
the polar cap models 
\citep{1975ApJ...196...51R,1979ApJ...231..854A,2000ApJ...532.1150Z,2002ApJ...568..862H},
the outer gap models 
\citep*{1973Natur.246....6H,1986ApJ...300..500C,1986ApJ...300..522C,1996ApJ...470..469R,1997ApJ...487..370Z,1999MNRAS.308...54H,2006MNRAS.366.1310T}
and the slot gap model
\citep*{2003ApJ...588..430M,2004ApJ...606.1143M}
have been developed to explain the observed pulsed light curve and
spectrum. 
The critical issue for these models is how the electric field along the
magnetic field, $E_{\parallel}$, is maintained locally although
the existence of the gap is implicit in these models.
Although the adequacy of these gap models is achieved statistically by
many further observations, but these models usually contain many
degrees of freedom, for example, which are the surface magnetic field
intensity of the pulsar, the inclination angle
between magnetic axis and rotational axis, the viewing angle from the
gaps and so on.
Recently, the radiation from the pulsar magnetosphere with
high-energy $\gamma$-ray bands has been observed by the Fermi
$\gamma$-ray telescope, and the $38$ gamma-ray pulsars have been
discovered 
\citep{2009Sci...325..840A,2009Sci...325..848A,2010ApJS..187..460A},
including $21$ radio-loud and 17 radio-quiet.
These observations could discriminate between the emission
models located on the polar region or outer region, i.g., polar cap,
slot gap, and outer gap. 
The Fermi telescope has also measured the spectral properties above
$10\,\mathrm{GeV}$ with a better sensitivity than EGRET.
It was found that the spectral shape of $\gamma$-ray emissions from the
Vela pulsar is well fitted with a power low 
(photon index $\Gamma\sim 1.5$) plus exponential cut-off 
($E_{\text{cut}}\sim 3\,\mathrm{GeV}$) model.
The discovered exponential cut-off feature predicts that the emissions
from the outer magnetosphere \citep{2009ApJ...696.1084A}
are more favored than the polar cap region \citep{1996ApJ...458..278D},
which predicts a super exponential cut-off with the magnetic
pair-creation. 
Furthermore, the detection of the radiation above $25\,\mathrm{GeV}$ bands
associated with the Crab pulsar has also predicted the high-energy
emission in the outer magnetosphere \citep{2008Sci...322.1221A} with
MAGIC (Major Atmospheric Gamma Imaging Cherenkov) telescope.

In this paper, we develop our global model 
such as linking the outer gap with the pulsar wind.
In order to construct a solution, we utilize particle simulation
in which particle motion and electric field are alternatively solved 
in the same way as the particle-in-cell (PIC) methods but with the magnetic
field given with dipole in our present model.
We simulate the generation of the plasma by pair creation
in the gap and then obtain a steady state of the axisymmetric
magnetosphere by solving the equation of motion including drag force due
to curvature radiation and the electromagnetic fields.
Since the motion of the individual particles is tracked, 
we can simulate any kind of drift motions for plasma, arising
from the radiation drag force, the centrifugal force and the gradient of
the magnetic field;
$\boldsymbol{f}_{\text{ext}}\times \boldsymbol{B}$ drift,
$\boldsymbol{E}\times \boldsymbol{B}$ drift and magnetic gradient drift.
Here, we provide results of additional calculations and
the detailed analysis for our steady solutions.
We shall show that the trans-field drift by radiation drag plays an
important role and eventually the electric field
dominant region expanding to both sides of the equatorial plane makes
a global poloidal current loop.
But this paper considers cases of low rate of pair creation, and
therefore modification of the original dipole field would be ignored as
well as in our previous work.
At first, the previous result (WS) of our steady solution was
reconstructed with $10$ times number of the particles and models with
several pair creation rates were carried out.
Our methods of calculations for the particle simulation are given in
section \ref{sec:jyzpY95D}, and the result is presented in section
\ref{sec:qJY51uua}, and section \ref{sec:E4dI9gcs} is for discussion.

\section{Numerical Method}
\label{sec:jyzpY95D}
\subsection{Outline}
\label{subsec:HzqK01wu}
We solve the motion of the particles and the electromagnetic
field alternatively so as to obtain a self-consistent steady structure
of the electromagnetic field and particle distribution.
This method is similar to the well-established particle-in-cell (PIC)
method.
However, to obtain a steady solution, we use static
solutions of the Maxwell equations, i.e., we  omit the effects of
time variation of the field (
$\partial \boldsymbol{E}/\partial t=\partial \boldsymbol{B}/\partial t=0$).
This enable us to use the programmable special purpose computer for
astronomical N-body problem; GRAPE-DR \citep{Makino2007}, which
calculates Coulomb interaction very rapidly with a high degree of
precision at the position of the plasma.

In our simulation, generation of the particle is considered, which is
not common in PIC methods.
There are two cases: (1) electrons or protons on the stellar surface
pop into the magnetosphere due to the unipolar induced electric field of
the star, (2) electron-positron pairs are created due to photon-photon
collision or magnetic pair creation in the magnetosphere. 
For the simplicity, the mass of both signed simulation particles is
taken to be the same value, that is, the proton and positron are not
distinguished, and the type of pair creation is not distinguished, that
is, the detailed process of pair creation is not considered.
We assume the pulsar to be a rotating spherical conductor in which
magnetic field is uniform with the magnetized axis being parallel to the
rotation axis.
In this paper, because we consider the case of low rate of pair
creation, which has much less current density than GJ model inside light
cylinder, the modification of the dipole magnetic field near the star by
the magneto-spheric current would be trivial.
For this reason, the magnetic field outside of the star is assumed to be
dipole, i.e.
\begin{align}
 \boldsymbol{B}=
 \frac{\mu}{r^3}
 (2\cos \theta \boldsymbol{e}_r+
 \sin \theta \boldsymbol{e}_{\theta})\text{,}
 \label{eq:fCOjl97b}
\end{align}
where $\theta$ is the colatitude, $r$ is the distance from the center of
the star, $\mu=B_0R^3/2$ is the dipole magnetic moment, $B_0$ is
magnetic field intensity at the poles, $R$ is the stellar radius and
$\boldsymbol{e}_r$ and $\boldsymbol{e}_{\theta}$ are the unit vector
along $r$ and $\theta$ directions, respectively.
These assumption gives the electric potential on the stellar surface,
\begin{align}
 \phi (\boldsymbol{R})=
 \frac{\Omega \mu \sin ^2\theta}{cR}+\text{constant,}
 \label{eq:h8aa4cVX}
\end{align}
where $\Omega$ is the angular velocity of the star.
This provides the inner boundary condition for the Poisson equation for
the electric potential.

If the outside of the star is a vacuum, the electrostatic potential
given by the solution of Laplace equation $\nabla^2\phi=0$ is
\begin{align} 
 \phi _{\text{v}}=
 \phi _{1}+\phi _{4}\text{,}
 \label{eq:z7TM3bvt}
\end{align} 
where
\begin{align} 
 \phi _{1}&=
 \frac{Q_{\text{sys}}}{r}\text{,}
 \label{eq:c5AMoi3x}\\
 \phi _{4}&=
 -\frac{\mu \Omega R^2}{3cr^3}(3\cos ^2\theta -1)\text{,}
 \label{eq:ikn86BHx}
\end{align} 
and $Q_{\text{sys}}=\alpha 2\mu \Omega/(3c)$, where $\alpha$ is the
non-dimensional net charge of the system.
If we follow the Jackson's Gedanken experiment 
(\citealt{1976ApJ...206..831J}),
we choose $\alpha =1$ as an initial condition, which corresponds to
the $+10$ model for KM.
Consequently, the vacuum electric field becomes
\begin{align}
 \boldsymbol{E}_{\text{v}}(\boldsymbol{r})&= 
 -\nabla \phi_{\text{v}}=
 \boldsymbol{E}_{4}(\boldsymbol{r})+\boldsymbol{E}_{1}(\boldsymbol{r})
 \text{,}
 \label{eq:5ISlnql1}
\end{align}
where
\begin{align} 
 \boldsymbol{E}_{4}(\boldsymbol{r})&=
 -\frac{\mu \Omega R^2}{cr^4}(3\cos ^2\theta -1)\boldsymbol{e}_r-
 \frac{\mu \Omega R^2\sin 2\theta}{cr^4}
 \boldsymbol{e}_{\theta}
 \text{,}
 \label{eq:nfN05Zyn}\\
 \boldsymbol{E}_{1}(\boldsymbol{r})&=
 \frac{Q_{\text{sys}}}{r^2}\boldsymbol{e}_r
 \text{.}
 \label{eq:Sxe0uB4n}
\end{align} 

This vacuum solution has the surface charge density
where
\begin{align}
 \sigma_{\text{v}}&=
 \frac{\mu\Omega}{4\pi cR^2}(3-5\cos ^2\theta )+
 \frac{Q_{\text{sys}}}{4\pi R^2}
 \text{.}
 \label{eq:16bKIhqc}
\end{align} 

Given $B_0$, $R$ and the light speed $c$, all quantities in our
simulation are normalized.
In the following, the barred quantities indicate that they are normalized
values.
For example, electric charge is measured in units of $B_0R^2$, and
$\bar{Q}_{\text{sys}}=Q_{\text{sys}}/B_0R^2$.
In the following subsections, we describe the detailed method of our
simulation.

\subsection{Method of Solution for the Electric field}
\label{subsec:avt7KD1g}
The space charge density in the magnetosphere is represented
by limited number of simulation particles, i.e.,
\begin{align}
 \rho _{\text{m}}(\boldsymbol{r})=
 \sum_{i=1}^nq_i\delta (\boldsymbol{r}-\boldsymbol{r}_i)
 \text{.}
\end{align}
In simulation, we are concerned with the super particles which have the
same mass and opposite signed charge with the same absolute value of the
charge and therefore the difference in the mass of ion and that of
positron is not considered.
As we introduce later, the normalized value of the mass and charge for
the super particle is represented by $\bar{m}$ and $\bar{q}$, respectively.
The electric potential in the magnetosphere is determined by
Poisson's equation, i.e., the solution for electric potential is given by
the superposition of the vacuum component \eqref{eq:z7TM3bvt} and space
charge component, which is calculate by 
$-\nabla^2 \phi_{\text{m}} = \rho_{\text{m}}$, with the boundary condition
$\phi_{\text{m}}(\boldsymbol{R})=0$, and we have
\begin{align}
 \phi_{\text{m}}(\boldsymbol{r})=
 \sum_{i=1}^nq_i
 \left[
 \frac{1}{|\boldsymbol{r}-\boldsymbol{r}_i|}
 -\frac{R/r_i}{|\boldsymbol{r}-(R/r_i)^2\boldsymbol{r}_i|}
 -\left(1-\frac{R}{r_i}\right)\frac{1}{r}
 \right]
 \label{eq:Dyer9Fj3}
\end{align}
for the electric potential,
\begin{align} 
 \boldsymbol{E}_{\text{m}}(\boldsymbol{r})=
 \sum_{i=1}^nq_i
 \left[
 \frac{\boldsymbol{r}-\boldsymbol{r}_i}
 {|\boldsymbol{r}-\boldsymbol{r}_i|^3}
 -\frac{R}{r_i}
 \frac{\boldsymbol{r}-(R/r_i)^2\boldsymbol{r}_i}
 {|\boldsymbol{r}-(R/r_i)^2\boldsymbol{r}_i|^3}-
 \left(1-\frac{R}{r_i}\right)
 \frac{\boldsymbol{r}}{r^3}
 \right]
 \label{eq:Uru09Aaa}
\end{align} 
for the electric field, and
\begin{align} 
 \sigma_{\text{m}}&=
  \sum_{i=1}^nq_i
 \left[
 \frac{1}{4\pi R}
 \frac{R^2-r_i^2}{|\boldsymbol{R}-\boldsymbol{r}_i|^3}-\frac{1}{4\pi R^2}
 \left(1-\frac{R}{r_i}\right)
 \right]
 \text{,}
 \label{eq:01laaIPd}
\end{align}
for the surface charge density.
Thus, the solution of the electric field with the boundary condition
\eqref{eq:h8aa4cVX}
is given in the forms,
\begin{align}
 \phi &=
 \phi_{\text{m}}+\phi_{\text{v}}
 \text{,}
 \label{eq:Zt7ymCq4}\\
 \boldsymbol{E} &=
 \boldsymbol{E}_{\text{m}}+\boldsymbol{E}_{\text{v}}
 \text{,}
 \label{eq:YNm35ydy}\\
 \sigma &=\sigma_{\text{m}}+\sigma_{\text{v}}
 \text{.}
 \label{eq:8tQ4anyC}
\end{align}

\subsection{Popping of particles from the star surface}
\label{subsec:Y70gkNjj}
GJ pointed out that
the rotation-induced electric field pulls charged particles 
from the stellar surface against the gravitational force.
In the following, we describe how this process is realized 
in our simulation.

In vacuum conditions, the scalar product of the electric field and the
magnetic field on the stellar surface is given by
\begin{align} 
 \boldsymbol{E}\cdot \boldsymbol{B}=
 \frac{\Omega RB_0^2}{c}\cos \theta 
 \left(\frac{\alpha}{3}-\cos ^2\theta \right)\text{.}
 \label{eq:e9CmvOy7}
\end{align} 
The sign of the electric field along the magnetic field
($E_{\parallel}$) changes on the surfaces 
$\cos \theta=\pm \sqrt{\alpha /3}$, i.e., the negative charged particles
pop out from the polar region and positive charged particles pop out
from the equatorial region.
Once the charge is emitted from the stellar surface, they experience 
$E_{\parallel}$ which changes sign on
$r=R+\sqrt{3/\alpha}\cos \theta$ for electron or on $\theta=\pi/2$ for
positron and ion (See, \citealt{1976ApJ...206..831J}).
The electrons are emitted from the lower colatitude region are
accelerated outward along the magnetic field line at first.
For electron, after passing over the surface at which the minimum of the
potential energy is given, the field aligned electric field
decelerate the particles and they are reflected toward the stellar
surface again.
Then they accumulate near the surface and makes the polar dome of
electrons because of the radiation drag force by the curvature radiation
for the particle.
At the same time, positrons or protons are
emitted from the higher colatitudes accumulating on the latter surface, i.e.,
the equatorial plane, and making the equatorial positively-charged
torus.
Thus, the vacuum electric field tends to be screened out by the
particles in the magnetosphere.

If the surface electric field along the magnetic field is shielded, the
surface charge density becomes,
\begin{align} 
 \sigma_{\text{GJ}}=\frac{3}{8\pi}\frac{B_0\Omega R}{c}\sin ^2\theta
 \text{.}
 \label{eq:z7cXDj3l}
\end{align} 
This surface charge density is resulted in the kink of the magnetic
field from the uniform one inside to dipole one outside under the ideal
MHD condition.
Although the surface charge density appearing on the stellar surface is
$\sigma$, only the excess charge $\sigma-\sigma_{\text{GJ}}$ is emitted
from the stellar surface and therefore we expect that the surface charge
density become $\sigma_{\text{GJ}}$ in the steady condition.
Thus, we replace $\sigma-\sigma_{\text{GJ}}$ by the movable simulation
particle.
For simplicity, the work function of the particles on the stellar
surface is not considered, that is the free emission of particles is
assumed.

\subsection{Parameter setting}
\label{subsec:75tsCbpN}
In our simulation, we use the super particle with artificially-enlarged
mass and charge of
$\bar{q}=10^{-5}, \bar{m}=10^{-10}$ or $\bar{q}=10^{-6}, \bar{m}=10^{-11}$.
The number of particles representing the charge 
$\rho_{\text{GJ,pole}}R^3$ is about $10^4$ particles for
$\bar{q}=10^{-5}$ model, which is same setting with WS, and $10^5$
particles for $\bar{q}=10^{-6}$ model, respectively.
Actual numbers of the particle is increased by pair creation, the total
number of particles in the simulation domain jumps tenfold for them
and therefore it needs heavy number of calculation related to square of
the number of particle for Coulomb interaction in one step.
Thus, we gain the privilege of using the GRAPE-DR, which is a
programmable special purpose computer for Astronomical N-body problems,
to calculate interaction between the particles.

In a realistic pulsar magnetosphere, the minimum scale of the motion of
the plasma is gyro radius, which is 
$r_{\text{g}}=1.7\times 10^{-2}r_3\gamma_7\,\mathrm{cm}$
for an electron or a positron, where $\gamma$ is Lorentz factor of
the particle $r_3=(r/R)^3$, and $\gamma_7=\gamma /10^7$ respectively.
We use super particles with a larger
mass-charge ratio than the real one.
We take $\bar{\Omega}=0.2$ and $\bar{m}/\bar{q}=10^{-5}$.
For the definition of the numerical parameters, a typical gyro radius
of the simulation particle on the pole is set to be 
$10^{-5}$ stellar radius, and therefore the corresponding time step is
set in $\mathrm{d}\bar{t}=10^{-5}$.
Since the valid time step of particle for the gyro motion increases
proportionately to the distance cubed from the star $\sim \bar{r}^3$, we
use the individual time step for each simulation particle adapted to
the position to reduce the total numbers of the integration time.
As we introduce the size of numerical domain later, we have to follow
these particles over several dynamical time length of the simulation
domain until a steady condition is achieved.
It takes about $6$ rotation periods of the star, i.e., 
$\bar{t}_{\text{sim}}\sim 200$.

For the unipolar induced electric potential by the rotational magnetized
conductor, the
available maximum energy of the particle is estimated by
the open field line voltage
\begin{align} 
 \phi_{\text{eff}}=\frac{\mu \Omega^2}{c^2}=
 1.6\times 10^{14}\left(\frac{P}{0.2\,\mathrm{sec}}\right)^{-2}
 \,\mathrm{Volt}
 \text{,}
 \label{eq:3VejUft4}
\end{align} 
which gives maximum Lorentz factor of
\begin{align} 
 \gamma_{\text{max}}=\frac{q\mu \Omega^2}{mc^4}=
 3.2\times 10^8\left(\frac{P}{0.2\,\mathrm{sec}}\right)^{-2}
 \label{eq:bN4uI8xa}
\end{align} 
for electron and positron.
Note that if the Lorentz factor of a particle becomes 
$\gamma_{\text{max}}$, the gyro radius in the vicinity of light cylinder
becomes the same order of the light radius. 
In other words, as expected,  the localized acceleration region has much
smaller potential drop than $\phi_{\text{eff}}$,
the gyro radius of the particles is much smaller than the light radius.
In our simulation, the gyro radius $\bar{r}_{\text{g}}$ of all
accelerated super particles is guaranteed smaller than the size of
the light radius; typically
$\bar{r}_{\text{g}}= 10^{-2}\bar{R}_{\text{l}}$ for an accelerated
particle corresponds to the Lorentz factor $0.1\gamma_{\text{max}}$ at
the light radius.

\subsection{Inner and Outer boundaries}
\label{subsec:5oJg9Vsd}
For the injection of the particles from the stellar surface, the surface
charge $(\sigma -\sigma_{\text{GJ}})\Delta S$ is replaced by the
simulation particles, where $\Delta S$ is the surface element of the
stellar surface.
If the particles are set on the stellar surface, they are pulled back
by the mirror charge, i.e., the second term of
\eqref{eq:Uru09Aaa}.
If this attractive field larger than the induced electric field, it
prevents extraction of particles from the stellar surface as an
artificial work function.
Then the assumption of free emission of the particle from the stellar
surface is not achieved.
We put the particles above altitude of
$\bar{h}_{\text{c}}$ from the stellar surface to avoid this artificial
attractive force.
We chose $\bar{h}_{\text{c}}=0.04$ for $\bar{q}=10^{-5}$ model,
$\bar{h}_{\text{c}}=0.013$ for $\bar{q}=10^{-6}$ model, respectively.
If the particles go back into the sphere with radius
$r=R+\bar{h}_{\text{c}}$, the particles are deleted in the numerical
region.
It is notable that the numerical thin vacuum layer does not change the
result at all.

To save the total number of simulation particles, we remove the
particles beyond a sphere with the radius
$R_{\text{OB}}$ and the removed charge is reduced from
$Q_{\text{sys}}$.
The size of outer boundary is taken to be larger than the one of the
accelerating region ($\sim R_{\text{l}}$), that is,
we chose $R_{\text{OB}}=10R_{\text{l}}$.
In our simulation, the removed particles beyond the outer boundary have
positive total energy (kinetic plus potential),
and we consider they would be outflow.
To examine the effect of the size of the outer boundary, the larger
outer boundary models, which are
$R_{\text{OB}}=20R_{\text{l}},\,30R_{\text{l}}$ are investigated, then
we confirmed that the structure of the magnetosphere inside the sphere
with radius $8R_{\text{l}}$ is not affected by the sizes of the outer
boundaries.
Thus, we use $R_{\text{OB}}=10R_{\text{l}}$ as our standard model.

\subsection{Particle motion}
\label{subsec:2wiAIhh8}
The equation of motion of the particles is given by
\begin{align} 
 m_ic\dD{\gamma_i\boldsymbol{\beta}_i}{t}=
 q_i\left[
 \boldsymbol{E}(\boldsymbol{r}_i)+
 \boldsymbol{\beta}_i\times \boldsymbol{B}(\boldsymbol{r}_i)
 \right]+
 \boldsymbol{f}_{\text{rad},i}
 \text{,}
 \label{eq:ElnOsz21}
\end{align} 
where $m_i$, $q_i$, $\boldsymbol{r}_i$ and $\boldsymbol{\beta}_i$  are
the mass, the charge, the position and the velocity of the $i$-th particle
and $\boldsymbol{f}_{\text{rad},i}$ indicates the drag force due to
curvature radiation of the accelerated particles.
In this paper, the magnetic field is not deformed by the magneto-spheric
current. The radiation drag force of the particles is given
\begin{align} 
 \boldsymbol{f}_{\text{rad},i}=
 -\frac{2}{3}\frac{q_i^2}{R_{\text{c}}^2}\gamma_i^4
 \frac{\boldsymbol{p}_i}{|\boldsymbol{p}_i|}\text{,}
 \label{eq:nmC9lg5I}
\end{align} 
where $R_{\text{c}}$ is the curvature radius along a dipole magnetic
field line at the position of the particle. 
Substituting $\varpi^2/r^3=1/r_{\text{eq}}=\text{const}$, the curvature
radius is given by
\begin{align} 
 R_{\text{c}}=
 f\frac{r}{3\sin \theta}\,\,\,\text{with}\,\,\,f=
 \frac{(4-3r/r_{\text{eq}})^{3/2}}{2-r/r_{\text{eq}}}\text{,}
 \label{eq:Cz4t1Agf}
\end{align} 
where $r_{\text{eq}}$ is axial distance of the point at the
intersection of dipole magnetic field line and the equatorial plane.
Because $1<f<4$ with $0<r/r_{\text{eq}}<1$, we take
$R_{\text{c}}=4r/(3\sin \theta)$ approximately.

If the Lorentz factor of the particle increases, \eqref{eq:nmC9lg5I}
become comparable to the Lorentz force, 
and such particle has 
$\boldsymbol{f}_{\text{rad}}\times \boldsymbol{B}$ drift motion crossing
the dipole magnetic surface.
The critical Lorentz factor for this effect is given by
\begin{align} 
 \gamma_{\text{d}}=\left(\frac{3\pi \mu}{qcP}\right)^{1/4}\text{.}
 \label{eq:zC3a9xBs}
\end{align} 
For a realistic pulsar magnetosphere, it corresponds to
$\gamma_{\text{d}} =3.6\times 10^7P_{\text{0.2}}^{-1/4}(R_{\text{c}}/R_{\text{l}})^{1/2}$.
Considering the dependency of the distance for $R_{\text{c}}$,
$R_{\text{c}}\sim R_{\text{l}}$ in the vicinity of the light cylinder.
For electron, $\gamma_{\text{d}}/\gamma_{\text{max}}=0.11$
with $P=0.2\,\mathrm{sec}$ and $\mu =10^{30}\,\mathrm{gauss\,cm^3}$, 
and thus the particles accelerated with $11\%$ of the open field line
voltage suffer from the such drift motion.
On the other hand, for our simulation particles, 
$\gamma_{\text{d,sim}}/\gamma_{\text{max,sim}}=0.0055$ for
$\bar{q}=10^{-5}$ model and
$\gamma_{\text{d,sim}}/\gamma_{\text{max,sim}}=0.0098$ for 
$\bar{q}=10^{-6}$ model, respectively, i.e., once the kinetic energy of
particles become $1\%$ of $q\phi_{\text{max}}$, the radiation drag force
is comparable to the Lorentz force.
As a result, the radiation drag force for the super particle would be
exaggerated.
To perform radiation drag force properly, we introduce the reduction
factor $\eta$, namely $\gamma_i$ in \eqref{eq:nmC9lg5I} replaced by
$\eta \gamma_i$ and take $\eta=0.05$ for $\bar{q}=10^{-5}$ model and
$\eta=0.089$ for $\bar{q}=10^{-6}$ model respectively although we
simulate with $\eta =1$ in previous work. 

\subsection{A Test Run: Reproduction of quiet solution}
\label{subsec:KElnqt62}
\begin{figure}
 \begin{center}
  \includegraphics[width=8cm]{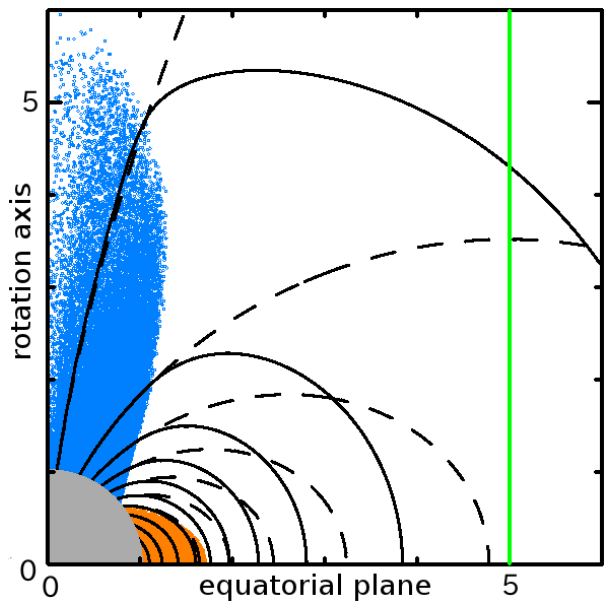} 
 \end{center}
 \vspace{-10pt}
 \caption{The particle distribution and the equipotentials for
 $\alpha=1$ and $q_{\text{unit}}=10^{-6}$ model.
 The orange and cyan points are
 positive and negative particles on the meridional plane
 ($\varpi ,z$), respectively. 
 The light cylinder is locates at $\varpi=5$ with the green line.
 The broken lines are the corotational equipotentials.
 The solid lines are the equipotentials. 
 The contour level is taken with equal interval at $0.5 \phi_{\text{eff}}$.}
 \label{fig:hn6HPer3}
\end{figure}

The pair creation in the gap has a significant role in the structure of
the magnetosphere.
If pair creation is suppressed, the static magnetosphere by 
KM is reproduced.
Following up to the Jackson's gedankenexperiment
\citep{1976ApJ...206..831J}, our simulation starts with the condition
that the positively charged magnetized rotating conductive sphere is
initially put at the origin in a vacuum.
The charge separated particles are emitted from the stellar surface and
are accelerated by the induced electric field along the magnetic field
lines.
But the energy of the particles is lost due to the radiation drag force
with time.
The particles are located at the bottom of the potential along the
magnetic field line, i.e. the force-free surfaces where the magnetic
field aligned electric field vanishes.
In vacuum, the force-free surfaces are polar domes 
 and the equatorial plane.
The inside of the polar force-free surfaces are filled with the negative
charges emitted from the polar caps of the stellar surface, and they
extend upward and form north and south force-free domes.
Similarly, the positive particles accumulate above and below the
equatorial plane and form a force-free torus. 
Eventually the field aligned electric field on the stellar surface is
shielded, and the static negative charged domes above
the poles and the static equatorial positive charged disc are formed.
Fig. \ref{fig:hn6HPer3} shows the static particle distribution with the
equipotentials $\phi$.
For comparison, they are superposed on the co-rotational equipotentials
which is given by $\phi_{\text{co}}=\Omega \mu \psi/c$, where 
$\psi=\mu \sin^2 \theta /r$ is the dipole magnetic flux function.
The equipotentials follow the co-rotational equipotentials in the
negative charged clouds and the inner part of the positive charged torus
close to the stellar surface.
In the cusp of the torus, the two kinds of equipotentials are deviated
but they are parallell.
Roughly, the $E_{\parallel}$ is screened out in the both signed charged
clouds.
Meanwhile, the electric field in the vacuum region faces the direction
such that particles draw back to the cloud with the same sign, and
therefore the cloud-vacuum boundaries are stable.

The force-free plasma in the cloud has toroidal velocity by 
$\boldsymbol{E}\times \boldsymbol{B}$ drift, which is given by
\begin{align} 
 v_{\text{t}}=\varpi \Omega +c\varpi \dD{\phi_{\text{nco}}}{\psi}\text{.}
 \label{eq:AItr35dd}
\end{align} 
where $\varpi$ is the radius in the cylindrical coordinates and 
$\phi_{\text{nco}} \equiv \phi-\phi_{\text{co}}$.
The coincidence of $\phi$ and $\phi_{\text{co}}$ in the cloud means
$\mathrm{d}\phi_{\text{nco}}/\mathrm{d}\psi=0$, i.e., the co-rotational
motion of the plasma.
The value of the potentials fall off faster than the co-rotational one
in the outer part of the disc, and therefore the term 
$c\varpi \mathrm{d}\phi_{\text{nco}}/\mathrm{d}\psi$ gives
super-rotation in the \eqref{eq:AItr35dd}.
In this structure,
the toroidal velocity
of the plasma in the outer part of the disc was less than light speed
because of the size of the disc is much less than the light cylinder.

\begin{figure}
 \begin{center}
  \includegraphics[width=8cm]{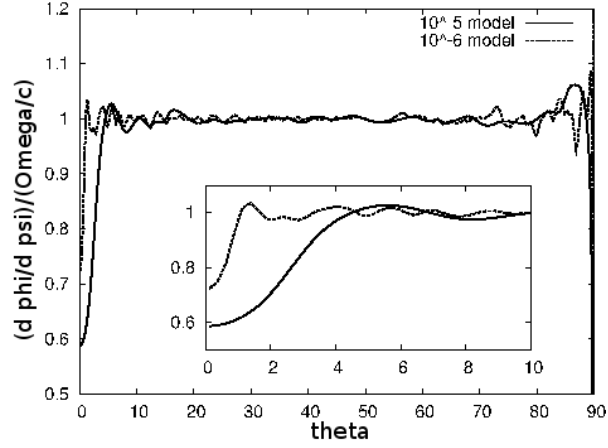}
 \end{center}
 \vspace{-10pt}
 \caption{$\mathrm{d}\phi/\mathrm{d}\psi$ nomalized by $\Omega/c$ versus
 the colatitude on the inner boundary. The small panel is close up near
 the polar region from $0\,\mathrm{deg}$ to $10\,\mathrm{deg}$.}
 \label{fig:g29EjyZj}
\end{figure}

As mentioned in subsection \ref{subsec:5oJg9Vsd}, note that our
simulation has artificial thin vacuum layer between the inner boundary
and the stellar surface.
The region causes deviation of the toroidal velocity from corotation in
the polar dome.
Fig. \ref{fig:g29EjyZj} shows $\mathrm{d}\phi/\mathrm{d}\psi$ for
the $10^{-5}$ model and the $10^{-6}$ model in unit of $\Omega/c$ on the
inner boundary.
This curve has a symmetry on both sides of the equator.
Most of the cloud co-rotates except for near the poles and the
equator.
The maximum deviation from the co-rotational velocity is $40
\%$ for the $10^{-5}$ model and $30 \%$ for the $10^{-6}$ model,
respectively.
Although the error is large on the equatorial plane, an affect near the
equator is trivial because the co-rotational velocity is originally
small at this place.
But the deviation by the numerical error near the pole is significant if
polar electric domes expand in the vicinity of the light cylinder.
In our simulation, the polar force-free electric dome, which connects
to the stellar surface along the magnetic field anchored at
$3\,\mathrm{deg}$ for $10^{-5}$ model and $1\,\mathrm{deg}$ for
$10^{-6}$ model, has about $10\%$ sub-rotational velocity from
the light speed with the distance of light radius.

\subsection{The treatment of pair creation and initial condition}
\label{subsec:EX1zuog9}
\begin{figure}
 \begin{center}
  \includegraphics[width=8cm]{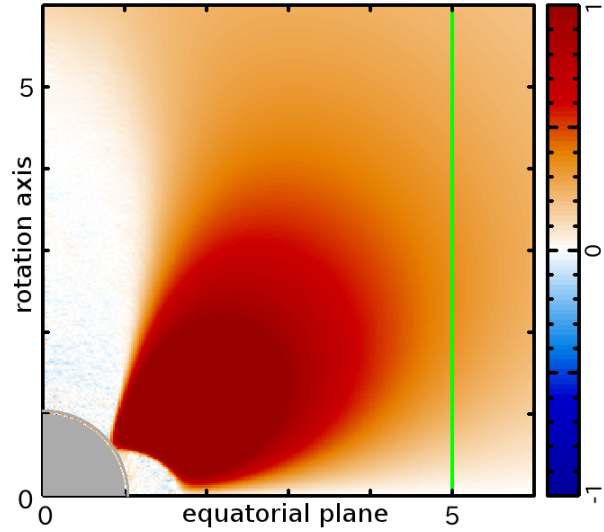}
 \end{center}
 \vspace{-10pt}
 \caption{The $E_{\parallel}$ intensity map for the pair starved static
 structure with $\bar{q}=10^{-6}$ model on the meridional plane, where
 the color intensity  is nomalized by $B_{\text{l}}$.
 The light cylinder locates at $\varpi=5$ with green line.}
\label{fig:e2caJC6c}
\end{figure}

Fig. \ref{fig:e2caJC6c} shows the distribution for strength of the
electric field along the magnetic field line ($E_{\parallel}$) on the
meridional plane for the static solution. 
A strong intensity  region is appeared in the middle latitudes, while
the less intensity is appeared in the charge clouds. 
For the static solution, there is a vacuum gap in the middle
latitude with a potential drop of $\sim 2\phi_{\text{eff}}$. 
In the vacuum gap, the maximum  intensity of $E_{\parallel}$ is
typically $E_{\parallel ,\text{max}}\sim 3B_{\text{l}}$ at
$(\varpi,z)=(1.8R,1.4R)$ on meridional plane, where $B_{\text{l}}\equiv
\mu /R_{\text{l}}^3$ is the dipole magnetic field intensity at the light
radius on the equator.
If the charge particles are injected into the gap, they will be
immediately  accelerated to ultra-relativistic regime by the
$E_{\parallel}$, and radiate the $\gamma$-rays by the curvature
radiation process, whose power is given by
\begin{align} 
 \dot{\epsilon}_{\text{c}}=\frac{2q^2\gamma}{3R_{\text{c}}^2}
 \text{.}
 \label{eq:0qrDL3hv}
\end{align} 
The typical Lorentz factor of the accelerated particles is estimated by
the force balance between the radiation drag force and the electric
force, i.e., $qE_{\parallel}=\dot{\epsilon}_{\text{c}}/c$, which gives
\begin{align} 
 \gamma_{\text{sat}}=
 \left(\frac{3\mu R_{\text{c}}^2}{2qR_{\text{l}}^3}\right)^{1/4}
 \text{.}
 \label{eq:Tn1wb1Vo}
\end{align} 
Then, the typical energy of the emitted $\gamma$-rays is given by
\begin{align} 
 \epsilon_{\gamma}=
 \frac{3hc\gamma_{\text{sat}}^3}{4\pi R_{\text{c}}}
 \text{.}
 \label{eq:t8xNuD4t}
\end{align} 
With typical parameters of the pulsars, we obtain the Lorentz factor 
of $\gamma_{\text{sat}}= 1.1\times 10^6P_{\text{0.2}}^{1/2}(R_{\text{c}}/R_{\text{l}})^{1/2}(E_{\parallel}/B_{\text{l}})^{1/4}$
and the photon energy of $\epsilon_{\gamma}=1.4\times P_{\text{0.2}}^{3/2}(R_{\text{c}}/R_{\text{l}})^{1/2}(E_{\parallel}/B_{\text{l}})^{3/4}\,\mathrm{GeV}$.
In the present work, we simplify the pair-creation process, that is, (1) we 
do not distinguish between the photon-photon pair-creation process and 
the magnetic pair-creation process, and (2) we ignore the effects of 
the collision angle between the $\gamma$-rays and soft-photons (or the 
magnetic field). 
In our simulation, we determine the pair-creation
position using the condition that $\epsilon_{\gamma}>2mc^2$, implying
the pair-creation is occurred at the position, where the electric field
is stronger than 
$E_{\text{cr}}=6.6\times 10^{-5}P_{\text{0.2}}^2(R_{\text{c}}/R_{\text{l}})^{2/3}B_{\text{l}}$.
$\epsilon_\gamma > 2mc^2$ is not applicable in reality because of the
effect of collision angle of the two photons.
On the other hand, we shall see below that actual setting of the value
of $E_{\text{cr}}$ is much larger than this value in the numerical
simulation since the accuraly of the elelctric field is much larger than
$E_{\text{cr}}$.

For our simulation, the pair creation is performed in the following way.
We introduce the critical electric field intensity to generate pairs in the
simulation as parameters, that is 
$\bar{E}_{\text{cr}}/\bar{B}_{\text{l}}=0.25, 0.125, 0.0625$ although
they are larger than the realistic value.
At first, the equally-spaced grid points $(\bar{r}_i, \theta_j)$ are
prepared with spherical coordinates in the meridional plane, where 
$\Delta \bar{r}=0.1$, $\Delta \theta=2^{\circ}$ for $\bar{q}=10^{-5}$
model and
$\Delta \bar{r}=0.1$, $\Delta \theta=1^{\circ}$ for $\bar{q}=10^{-6}$
model, respectively.
If the field-aligned electric field at a grid points
$\bar{E}_{\parallel}(i,j)$ is larger than $\bar{E}_{\text{cr}}$, then we
put on $n_{\pm}=n_{\text{M}}$ pairs around the grid, where $n_{\text{M}}$ is
multiplicity of the pair creation.
The generation of pairs is repeated in every
$\bar{\tau}_{\text{pr}}$, where we take $\bar{\tau}_{\text{pr}}=2.0, 1.0, 0.2$,
which are less than the light crossing time of the light radius.
We consider $n_{\text{M}}$ as a constant parameter and the lowest cases
$n_{\text{M}}=1$ are carried out to guarantee the pure dipole magnetic
field assumption.
We set up the static solution as an initial condition of the
simulation with pair creation effect, e.g., $3000$ pairs are generated in
the gaps with $\bar{q}=10^{-5}$ model with $n_{\text{M}}=1$ in initial.
Table \ref{table:8yrP4fiZ} provides the numerical set of parameters of
simulation.

Note that the perturbation of a simulation particle makes error
electric field.
The maximum error of electric field caused by a simulation particle is
roughly given by $E_{\text{err}}=\bar{q}/\Delta^2$, which is
$0.25B_{\text{l}}$ for $10^{-5}$ model and $0.025B_{\text{l}}$ for
$10^{-6}$ model in the present parameter setting, where $\Delta$ is
typical interval of the grid to estimate $E_{\parallel}$ and we choose
$\Delta=0.1R$.
If $E_{\text{cr}}$ is taken to be much smaller than $E_{\text{err}}$,
the pair is generated excessively and therefore the calculation is
immediately broken by running out of the limit of the number of
particles in our simulation.

\begin{table}
  \begin{tabular}{c|c|c|c|c|c} \hline
   name & $\bar{q}$ & $\bar{E}_{\text{cr}}$ &  $\tau_{\text{pr}}$ &
   $\Delta r$ & $\Delta \theta$ \\  \hline
   A0 & $10^{-5}$ & $10^{-3}$ &           $2.0$ & $0.1$ & $2^{\circ}$\\
   A1 & $10^{-5}$ & $10^{-3}$ &           $1.0$ & $0.1$ & $2^{\circ}$\\
   A2 & $10^{-5}$ & $10^{-3}$ &           $0.2$ & $0.1$ & $2^{\circ}$\\
   A3 & $10^{-5}$ & $10^{-3}$ &           $0.1$ & $0.1$ & $2^{\circ}$\\
   A4 & $10^{-5}$ & $5\times 10^{-4}$   & $2.0$ & $0.1$ & $2^{\circ}$\\
   A5 & $10^{-5}$ & $2.5\times 10^{-4}$ & $2.0$ & $0.1$ & $2^{\circ}$\\
   B0 & $10^{-6}$ & $10^{-3}$ &           $0.2$ & $0.1$ & $1^{\circ}$\\ \hline
  \end{tabular}
 \caption[]{\small{The numerical parameter of simulation.}}
  \label{table:8yrP4fiZ}
\end{table}

\section{Result}
\label{sec:qJY51uua}
\subsection{Quiet Solution: without pair creation}
\label{subsec:TOd7jw2b}
\begin{figure}
 \begin{minipage}[t]{.45\linewidth}
   \begin{center}
    \includegraphics[width=4cm]{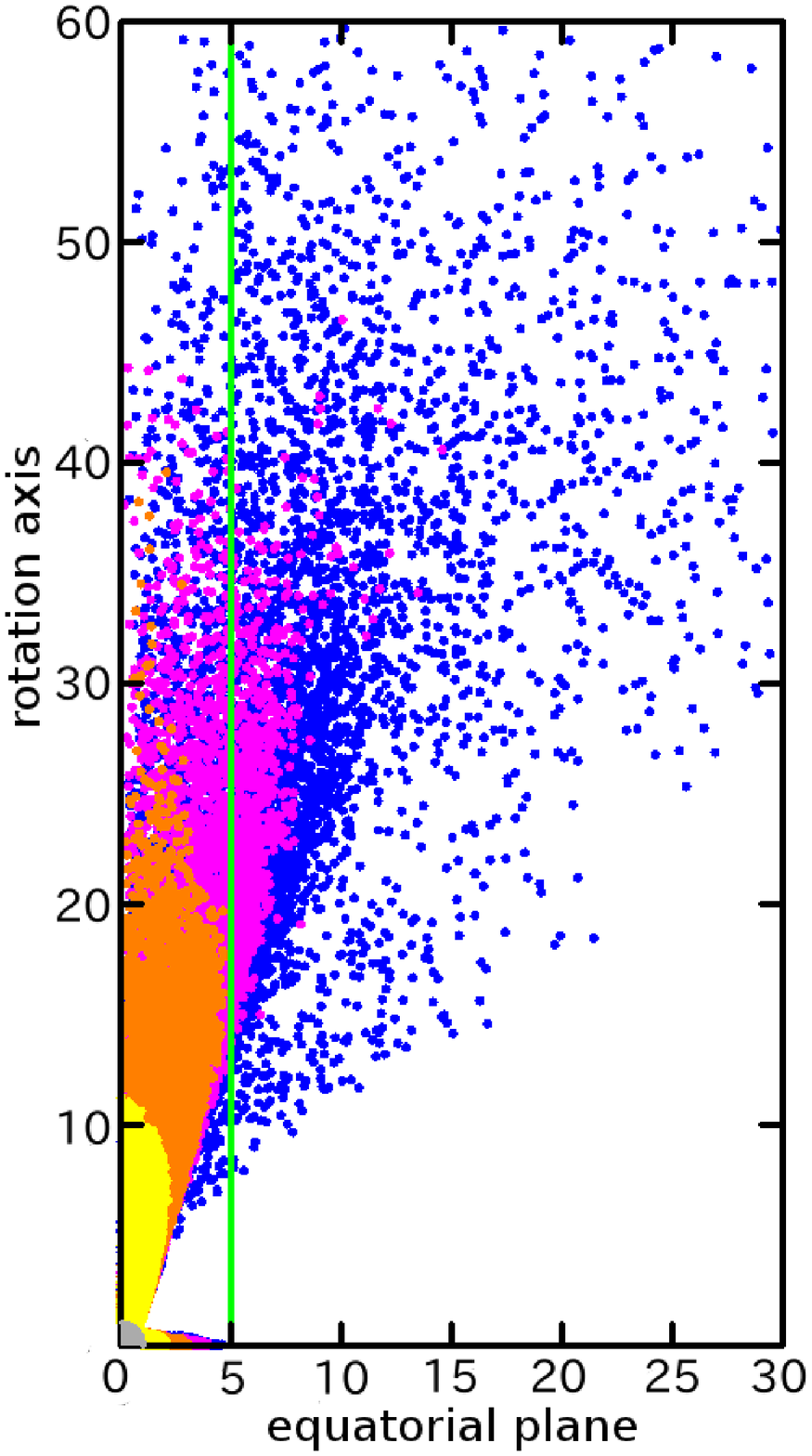}
   \end{center}
 \vspace{-10pt}
 \end{minipage}
 \hspace{.05\linewidth}
 \begin{minipage}[t]{.45\linewidth}
   \begin{center}
   \includegraphics[width=4cm]{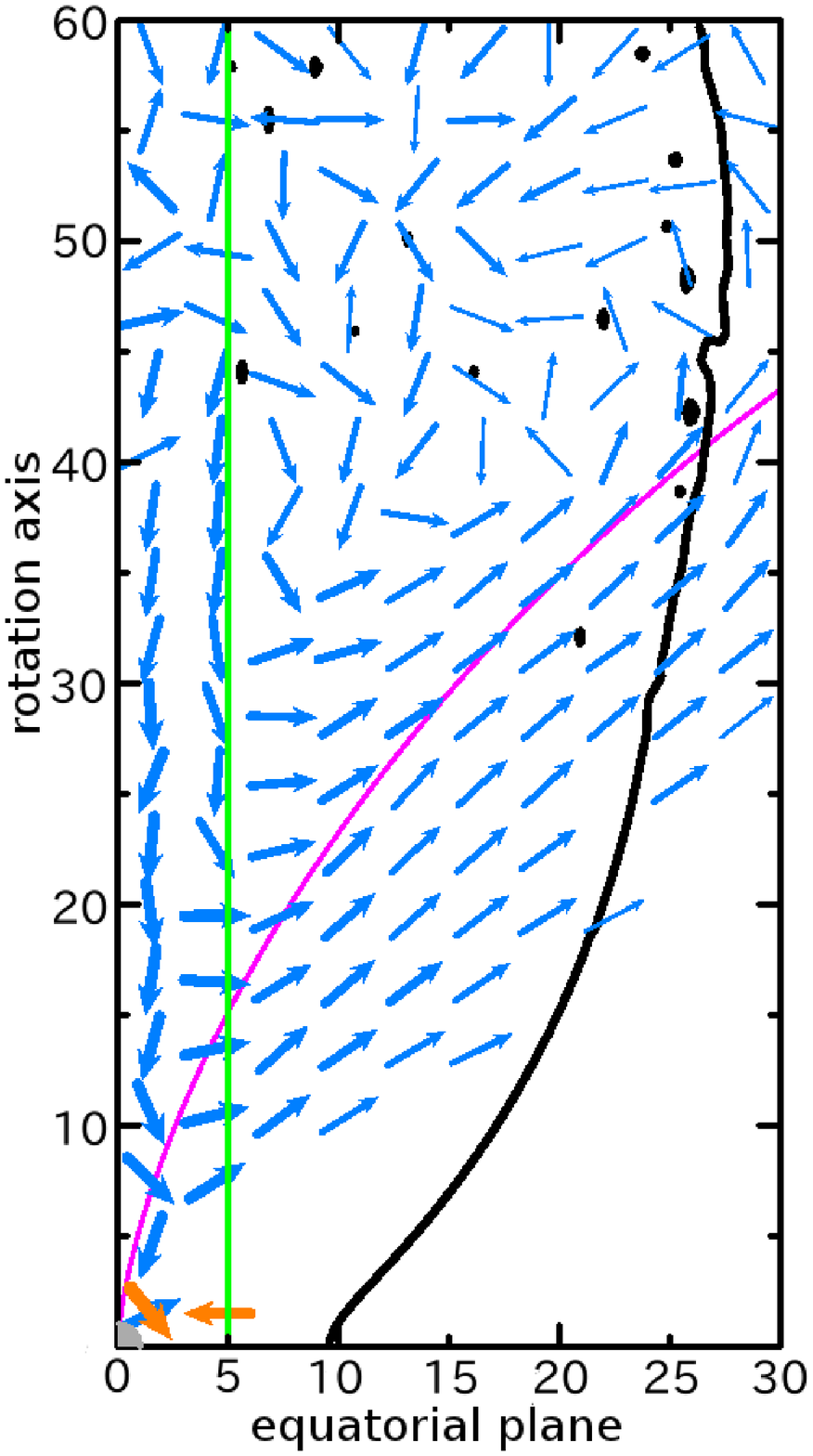}
   \end{center}
 \vspace{-10pt}
 \end{minipage}
 \caption{Left panel: The distributions of the particles on the meridional
 plane with simulated $Q_{\text{sys}}$'s, Colored dots indicate 
 $\alpha=0.75,  0.5, 0.4, 0.3$ with blue, magenta, orange and yellow
 respectively.
 The light radius is located at $\varpi=5$ with green line. 
 Right panel: the current density pattern on meridional plane for
 $\alpha=0.3$ model.
 The arrows indicate the direction of the current density of negative
 charged particle(blue) and positive charged particle(red).
 The magenta curve indicates magnetic field line footed on the surface
 with colatitude $4.5$ degree.
 The boundary surface with $E=B$ is drawn by solid curve.}
 \label{fig:3zFhm7fD}
\end{figure}

As has been shown, if pair creation is suppressed, the electrosphere is
composed of electronic domes above the pole, a positronic
or ionic equatorial disc and vacuum gaps in the middle latitudes.
It takes about one rotation period until the quiet state is
achieved with $\alpha=1$.
We also simulate for smaller system charges of 
$\alpha = 0.75, 0.5, 0.4, 0.3$ with $\bar{q}=10^{-6}$ model.
The smaller the net charge, the larger the extent of the polar dome (see
left panel of Fig. \ref{fig:3zFhm7fD}).
This tendency has been stated by \cite{1985A&A...144...72K} $+10$ and
$+4$ models which correspond to our $\alpha=1$ and $0.4$ models,
respectively.

The same structures are identified as well as KM with $\alpha\geq 0.4$
in our simulation because the polar dome falls in the light cylinder and
therefore it satisfies the condition of rigid constraint for the
particle by the dipole magnetic field.
But the model with $\alpha=0.3$, the static structure is changed.
The global flow pattern of the particles for the model with $\alpha=0.3$ is
shown in the right panel of Fig. \ref{fig:3zFhm7fD}.
When we choose a model with $\alpha<0.4$, the polar dome extends beyond
the light cylinder and the azimuthal velocity of the particle in the dome
increases with the axial distance although the azimuthal velocity is
slightly small from the corotational velocity.
Thus, the Lorentz factor increases close to $\gamma_{\text{d}}$ near the
light cylinder and then the radiation drag force of the particle makes
$\boldsymbol{f}_{\text{rad}}\times \boldsymbol{B}$ drift just out side
of the light cylinder.
Thus, the particles emitted along the polar magnetic field line migrate
into the inner magnetic surface if they pass through the light
cylinder, and they returns star just higher colatitude of their
departure point.
As a result, there are outflows of the negative particle 
from the polar region, which are bounded the magnetic surface footed on the
stellar surface with $4.5$ degree.
Note that, for our case, the whole charge cloud is still in magnetic
field dominant region ($B>E$).
When the outflows go beyond the light cylinder, they move across the
inner magnetic surface by the 
$\boldsymbol{f}_{\text{rad}}\times \boldsymbol{B}$ drift and they become
inflow by the quadrepole electric field in the vacuum region at the
middle latitudes.
Thus, the closed poloidal current loops above the pole are formed in
several light radius. 
This is more favorable rather than the faraway loop of the electron made
by attraction of the monopole electric field of the star suggested by
\cite{1976ApJ...206..831J}.
Our result confirmed the same flow pattern of electrons given by Rylov
(\citeyear{1977Ap&SS..51...59R}).
Although a part of the edge of the dome has fast azimuthal velocity,
the current density of the dome in the vicinity of the light cylinder is
typically $10^{-5} \rho_{\text{GJ,pole}}c$, and therefore the
modification of the dipole magnetic field should be ignored.

\begin{figure}
 \begin{center}
  \includegraphics[width=8cm]{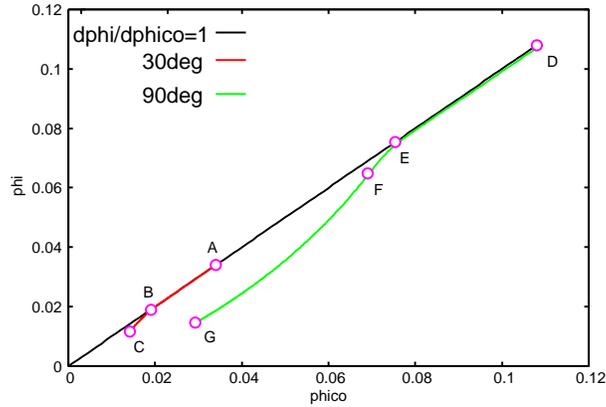}
 \end{center}
 \vspace{-10pt}
 \caption{The electrostatic potential $\phi$ along the radial line with
 $\theta =30^{\circ}$ (A-B-C) and with $\theta = 90^{\circ}$ (on the
 equatorial plane) (D-E-F-G) as a function of the magnetic stream
 function $\psi$.
 The solid line indicate the co-rotational potential
 $\phi_{\text{co}}$.}
 \label{fig:y0Umda6P}
\end{figure}

\begin{figure}
 \begin{center}
  \includegraphics[width=8cm]{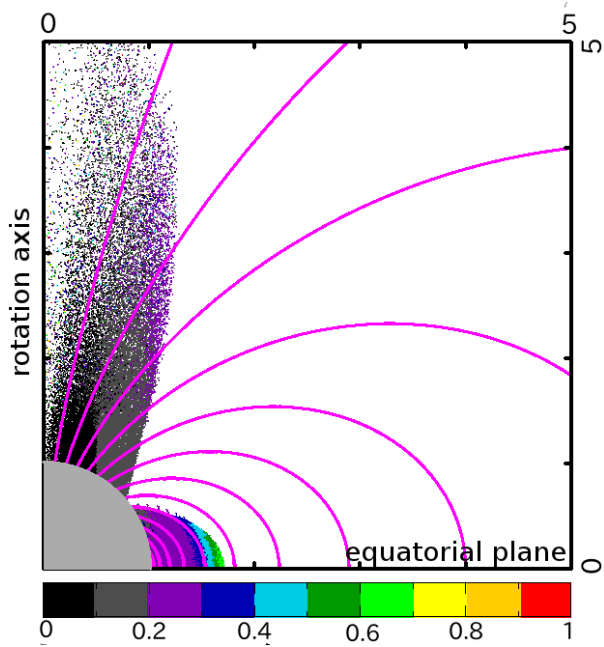}
 \end{center}
 \vspace{-10pt}
 \caption{The particles color-coded with toroidal velocity on the
 meridional plane. 
 The meaning of the colors is shown in the color bar just below the
 equatorial plane, where each color indicates the co-rotation speed at
 each axial distance.
 The solid curves are dipole magnetic field lines.}
 \label{fig:IKb78adz}
\end{figure}

Fig. \ref{fig:y0Umda6P} shows $\phi$ along two radial directions with
$\theta=30^{\circ}$ and $\theta=90^{\circ}$ (equatorial plane), respectively.
The former radial line passes through the polar dome, while the latter
does through the equatorial disc.
It is notable that the derivative $\mathrm{d}\phi/\mathrm{d}\psi$
indicates the angular velocity of the clouds.
Along the line with $\theta=30^{\circ}$, the electric potential
$\phi(\psi)$ follows the co-rotational $\phi_{\text{co}}$ in between
A and B in the figure, which correspond to the stellar surface and
the surface of the cloud (cloud-vacuume boundary).
This indicates that the cloud co-rotates with the star in the cloud. 
Along the line with $\theta= 90^{\circ}$ (equator), the inner part of
the disc between $D$ and $E$ follows the co-rotational
$\phi_{\text{co}}$, and therefore they show co-rotation and
$E_{\parallel}=0$.
The point $F$ corresponds to the top of the equatorial disc.
Because of $\mathrm{d}\phi/\mathrm{d}\psi >\Omega /c$ in between $E$
and $F$, the positive charge at the edge of the disc are in
super-corotation.
This part is connected to the vacuum gaps in the middle latitudes along
the magnetic field line.
The electric potential in the outer part of the disc decreases faster
than the co-rotational potential.
Thus, the super corotation of the disc top is obtained.
Outside of the clouds (between B and C) is in vacuum, and $\phi$
deviates from $\phi_{\text{co}}$ , i.e., $E_{\parallel}$ exists. 
These features can actually be seen with the velocity of the particle in
Fig.
\ref{fig:IKb78adz}, i.e., one can see the co-rotational motion in the
polar domes and in the inner part of the disc, the super-rotation in the
outer part of the disc.
The azimuthal velocities are not close to the light speed at the top of
the equatorial disc, and therefore the radial diffusion mechanism by
rotational inertia and radiation drag are negligible in such cases.
Outside of the disc (between $F$ and $G$) is in vacuum.

\subsection{Active Solution: with pair creation}
\label{subsec:Ued6jM2g}
The steady solutions are obtained in about 4 rotation periods.
For comprehending the trajectories of the particle, we have to know the
magnitude relation between electric field and magnetic field and the
effect of the drift motion caused by radiation drag force.
When the plasma is in the magnetic field dominant region (MDR), it tends
to move along magnetic surface.
However, once it is in the electric field dominant region (EDR),
particles are accelerated toward the direction of the electric field.
If $E_{\perp}\sim B$ in MDR, the velocity of the 
$\boldsymbol{E}\times \boldsymbol{B}$ drift of the plasma is close to
light speed.
Then the radiation drag force for the particle is comparable to the
Lorentz force, that is the Lorentz factor of the plasma becomes 
$\gamma_{\text{d}}$, which is given by \eqref{eq:t8xNuD4t}.
Such plasma drifts perpendicular to the magnetic
surface by the $\boldsymbol{f}_{\text{rad}}\times \boldsymbol{B}$ drift.
It is notable that the 
$\boldsymbol{f}_{\text{rad}}\times \boldsymbol{B}$ drift is opposite
direction depending the sign of the charge, i.e., the positive charged
particle drifts outward and the negative charged particle drifts inward.

\begin{figure}
 \begin{center}
  \includegraphics[width=8cm]{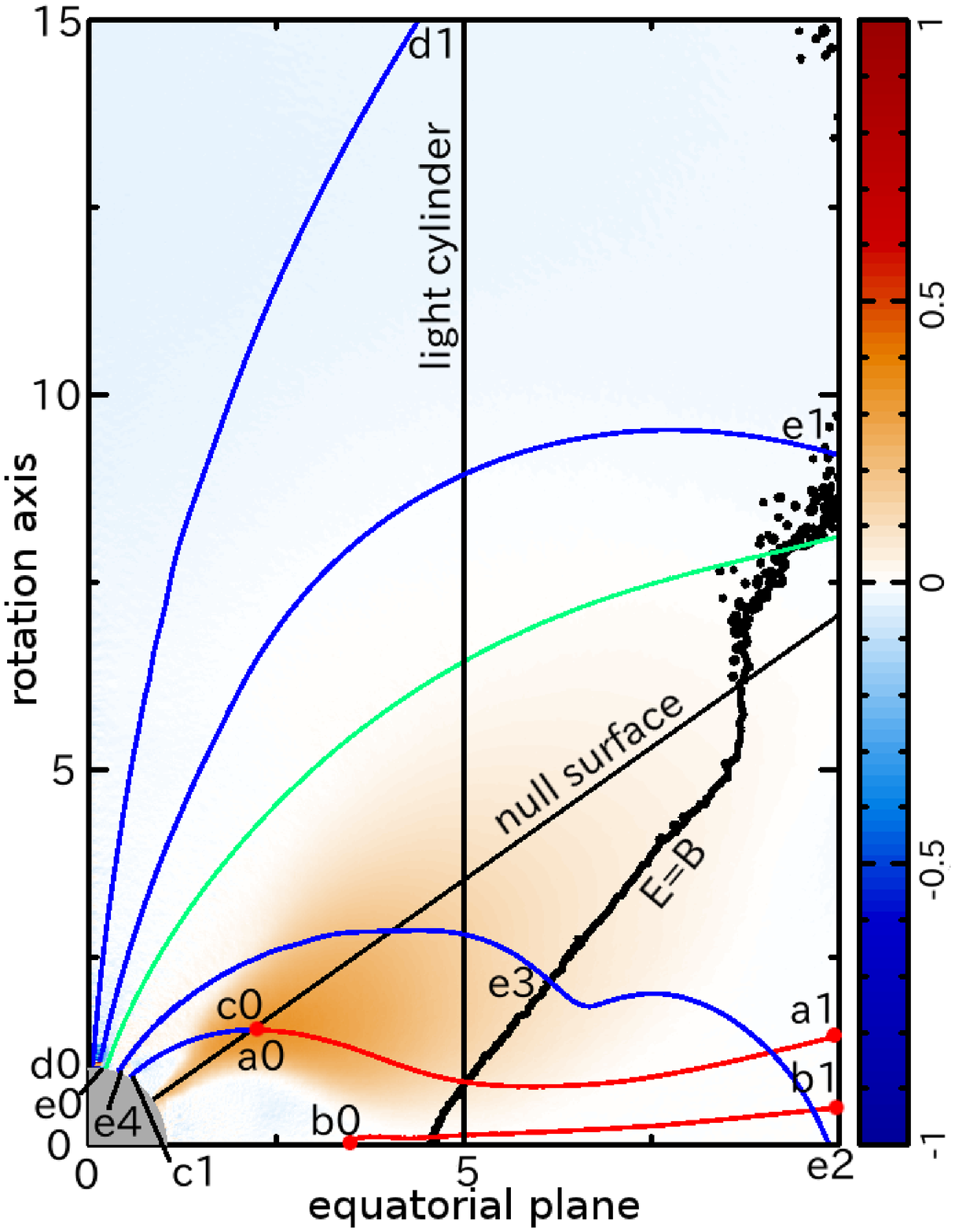}
  \includegraphics[width=8cm]{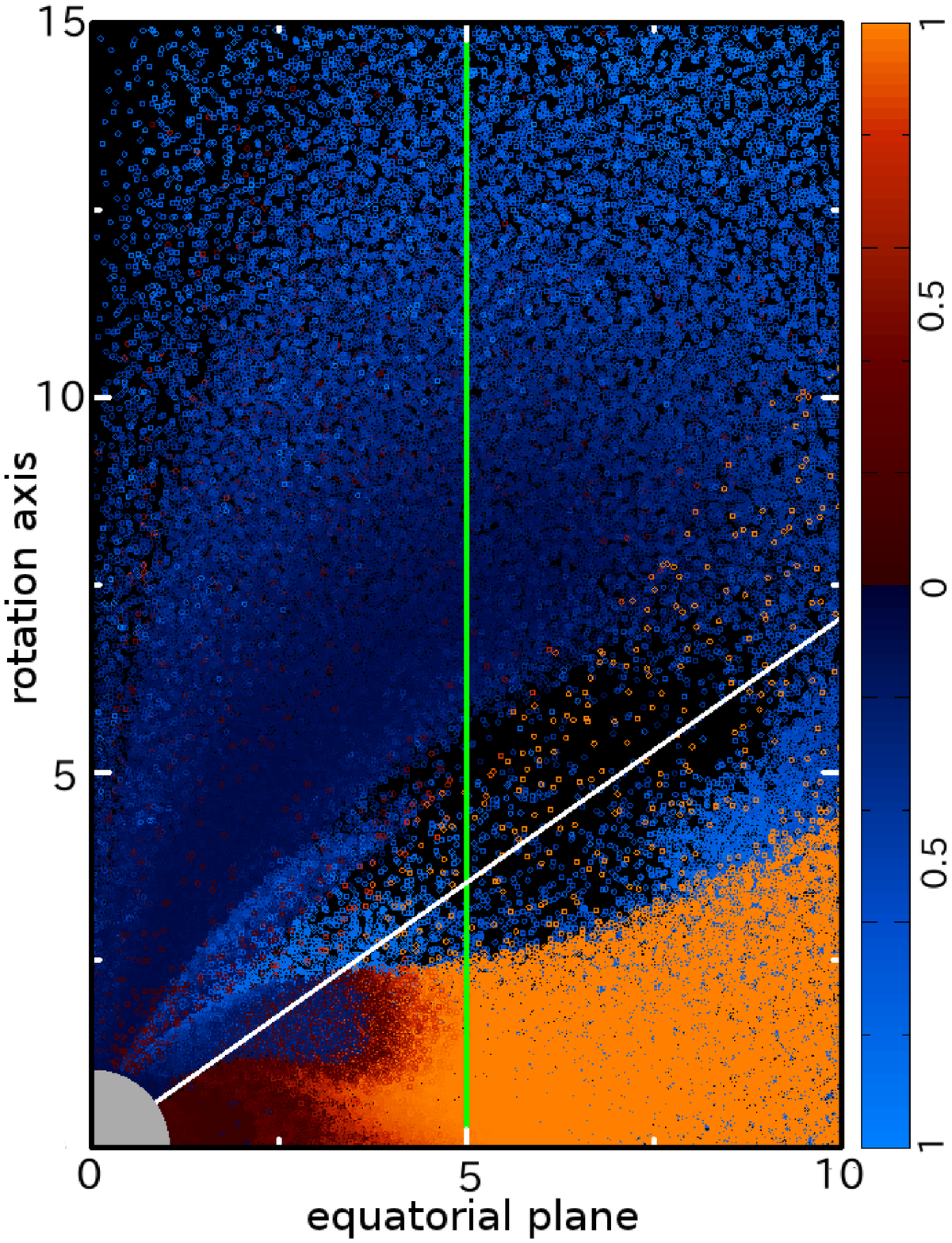}
 \end{center}
 \vspace{-10pt}
 \caption[]{\small{Left panel: typical trajectories of the particle and
 $E_{\parallel}$ intensity map, the curves are color-coded with blue
 (for the negative charged particle), red (for the positive charged
 particle) and the color map indicates electric field intensity
 normalized by $B_{\text{l}}$. 
 The thick solid curve indicates isosurface with $E=B$, and the solid
 curve with green means equipotential surface with $\phi =0$.
 Right panel: Lorentz factor of the particles, which are color-coded
 with the Lorentz factor normalized by $\gamma_{\text{d}}$. For
 both panels, the light cylinder is located at $\varpi =5$ and the line
 in the middle latitude is null surface. )}}
 \label{fig:Py9yk6wA}
\end{figure}

In the left panel of Fig. \ref{fig:Py9yk6wA}, the typical trajectories
of particle on the meridional plane are drawn with red and blue curves
and EDR and MDR are bounded with thick curves.
The EDR appears around the equatorial plane in a wedge-like shape beyond
$\varpi \sim 4.5=0.9R_{\text{l}}$ with the opening angle of about
$50^{\circ}$ from the equatorial plane.
Although the EDR appears due to the monopole electric field in the quiet
model \citep{1976ApJ...206..831J,1977Ap&SS..51...59R},
our present model has almost no net charge.
The EDR seemed to have been formed by the global structure of the charge
clould, and in particular it would be due to the growing equatorial
positive charged disc.
The similar shape of EDR is discussed in the force-free model around
Y-point by Uzdensky (\citeyear{2003ApJ...598..446U}).
The angle between the EDR and the equatorial plane in our result is just
wider than Their EDR.
It is interesting that the structure of EDR is very similar although our
result precludes the force-free approximation and assumes the magnetic
field to be dipole elsewhere.
In the right panel of Fig. \ref{fig:Py9yk6wA}, the luminous color-coded
particle have the Lorentz factor being comparable with
$\gamma_{\text{d}}$, i.e., the radiation drag force for the particle is
comparable to the Lorentz force $qB_{\text{l}}$.
They are mainly in the vicinity of the light cylinder.
The positive charged particle near the equatorial plane just inside the
light cylinder has the Lorentz factor with $\gamma_{\text{d}}$.
At the same time, the negative charged particle in the vicinity of the
light cylinder with the height above $10R$.
The dusk color-coded dots in the right panel of Fig. \ref{fig:Py9yk6wA}
indicates the non-accelerated particles, i.e., they are co-rotating
equatorial disc and conically-shaped polar domes.
The background color of the left panel of Fig. \ref{fig:Py9yk6wA}
indicate the intensity of $E_{\parallel}$, which are color-coded red 
($E_{\parallel}>0$), blue ($E_{\parallel}<0$) and
white ($E_{\parallel}=0$), respectively.
Both sides of the outer gap, the $E_{\parallel}$ is shielded and
there is the co-rotating charged clouds.

In the left panel of Fig. \ref{fig:Py9yk6wA}, the flow of the
positive particle generated in the outer gap goes into EDR, it easily
goes out beyond the light cylinder ($a_{0}\rightarrow a_{1}$).
The plasma at the outer edge of the disc at $b_{0}$ has a fast
azimuthal velocity, and therefore they are slipped out by
$\boldsymbol{f}_{\text{rad}}\times \boldsymbol{B}$ drift
($b_{0}\rightarrow b_{1}$).
Meanwhile the flow of the negative particles generated in the outer gaps
moves back to the star ($c_{0}\rightarrow c_{1}$) with colatitude
$24^{\circ}<\theta <34^{\circ}$ on the stellar surface, 
and re-emitted from the polar region.
The polar flow is separated by the magnetic surface
footed on the stellar surface with colatitude $10^{\circ}$, where the
equipotential surface of $\phi=0$ connects up to the star.
For the lower colatitude region, it is outflow drawn by 
$d_0 \rightarrow d_1$.
For the other region, it is circulation on the meridional plane,
which is starting from the polar annulus with the colatitude
$10^{\circ}< \theta <12^{\circ}$, and returning to the annulus 
with colatitude $18^{\circ}< \theta <24^{\circ}$ 
($e_{0}\rightarrow e_{1}\rightarrow e_{2}\rightarrow e_{3}\rightarrow e_{4}$).
On the way through $e_{0}\rightarrow e_{1}$, the flow having fast
azimuthal velocity migrates to the inner magnetic surface because of
$\boldsymbol{f}_{\text{rad}}\times \boldsymbol{B}$ drift.
Once it moves in EDR, it is accelerated and passes over the equatorial
plane. 
Because of the plane-symmetric fashion about the equatorial plane, the
flow having the same trajectory in the opposite hemisphere returns from
$e_2$, then it moves in MDR ($e_2\rightarrow e_3$).
Thus, it returns along a magnetic field line just outside
the pair creating region in the middle latitude 
($e_{3}\rightarrow e_{4}$).

\begin{figure}
  \includegraphics[width=7cm]{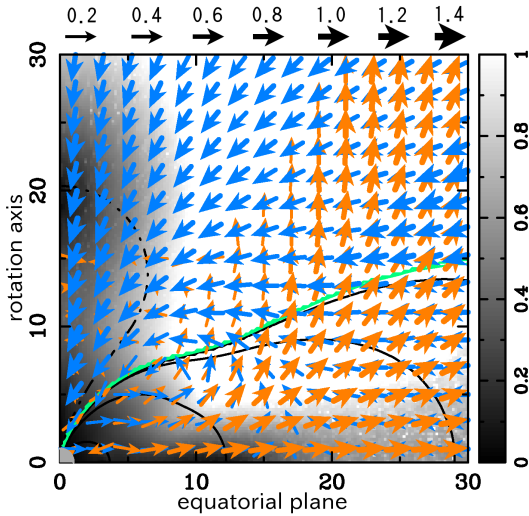}
 \caption[]{\small{The arrows indicate the current density made of
 positive charge (orange) and negative charge (cyan), which are
 normalized by $\Omega B/(2\pi)$.
 The intensity ratio of dipole magnetic field and the magnetic field
 made of the magnetospheric current
 is drawn by gray scale color-coded map, that is $|B_q|/|B_{\text{d}}|$.
 The curves are electrostatic equipotentials normalized by
 $\phi_{\text{max}}$  with common logarithm, where $\phi=0$ is colored
 with  green, the solid curves indicates $\phi>0$ and the dashed curve
 indicates $\phi<0$.}}
 \label{fig:s7rFZy0o}
\end{figure}

The overhead view of the steady solution is shown in
Fig. \ref{fig:s7rFZy0o}. 
The intensity ratio of the dipole magnetic field and the magnetic field
made of the magnetospheric current in the steady state is drawn with
gray scale map.
The current density normalized by $\Omega B/(2\pi)$ on the meridian
plane is drawn with color-coded arrows and the electrostatic
equipotentials normalized by the $\phi_{\text{eff}}$ are drawn with the
curves.
The green curves indicates that the value of the equipotential is zero.
The value of the potential is negative in the lower colatitude region
and is positive in the higher colatitude region.
The contour level is incremented by $-3,-2,-1,0$ with common logarithm
from the zero-equipotential surface for both signs of potentials.
It shows the assumption of dipole magnetic field is valid in the light
cylinder and especially round about the equatorial plane, although the
magnetic field made by the magnetospheric current is comparable to the
dipole field in the middle latitude about beyond $3R_{\text{l}}$, where
magnetic flux is modified to be opened.
The circulation pattern of the flow in the present result would turn
down if the modification of the magnetic field is concerned.
The magnetic field made of the magnetospheric current is trivial inside
the light cylinder, namely the structure of the outer gaps and pair
creation rate should not be affected in present model if the
modification of the dipole magnetic field is considered.
In contrast, outside of several light radius, the magnetic field made of
the magnetospheric current is comparable to the dipole magnetic field. 
However, there are almost in EDR, and therefore the particle flow is
controled by the electric field.
Concerning the global structure of the current and equipotential
surfaces, the negative charged flowed out in a lower latitude and the
positive charged flowed out in a higher latitude.
However, the system of the charge of our result is almost neutral, that
is the monopole electric field would not prevent reversed sign outflow at a
great distance.
Eventually both outflows have enough kinetic energy larger than the
potential energy at the outer boundary, and therefore they can reach
infinity beyond the outer boundary.
Inside of the sphere about with radius $3R_{\text{l}}$, there are poloidal
current loops caused by 
$\boldsymbol{f}_{\text{rad}}\times \boldsymbol{B}$ drift from equatorial
plane to the pole, which mainly consist of negative charged flow.
In previous paper, the parameter of the
radiation drag force, $\eta$ is taken to be unit so that the circulation
of the both signed charge caused by the 
$\boldsymbol{f}_{\text{rad}}\times \boldsymbol{B}$ drift is incident.
In present result, there are few returning positive charges although the
poloidal current structure is similar to that in previous work.

\section{Discussion}
\label{sec:E4dI9gcs}
\subsection{Comparison with the quiet solution}
\label{subsec:7YN8ujms}
\cite*{2001MNRAS.322..209S} concluded that their result for the axisymmetric
magnetosphere is quiet because the pair plasma which moves along the
dipole magnetic field lines screens out the $E_{\parallel}$ in the outer
gaps and therefore the pair creation is stopped. 
They implied that activity of pulsar is essentially caused by obliqueness
(see also \citealt{2004AdSpR..33..542M}).
Although the similar screening of the outer gaps is shown in our
simulation, but intensity of the electric field is maintained to
generate pair plasma.
We note that the our active solution arise from taking smaller
threshold intensity of electric field for pair creation than
\cite*{2001MNRAS.322..209S} and solving the equation of motion for each
particle without the assumption of restraint of plasma along the dipole
magnetic field line.
As a result, we showed that if the radiation drag of the particle and
the supply of the pair plasma in the gaps are performed, the quiet system
should be broken and migrates over to the active system with gaps and
both signed outflows of the plasmas nevertheless the rate of the pair
creation is taken to be small as the modification of the dipole magnetic
field is omitted.
Some fraction of the positive charge generated in the outer gaps
accumulates the equatorial plane, and therefore the super rotating disc
is growing and the azimuthal velocity increases, and then the Lorentz
factor increases as the light cylinder reaches in some fraction of 
$\gamma_{\text{d}}$.
The particles at the cusp of the disc start to leak out due to the 
$\boldsymbol{f}_{\text{rad}}\times \boldsymbol{B}$ drift.
If the potential drop in the outer gap is $f$ in fraction of
the effective voltage, the trans-field drift would take place 
beyond $(1-f)R_l$, i.e., it is likely that radiation drag causes
trans-field drift motion within the light cylinder.
As has been shown, this effect creates an outflow of positive charged
particles even if the dipole magnetic field is closed, and it is
favorable to keep the outer gaps.

As diocotron instability of differential rotation disc is pointed by
\cite*{2004IAUS..218..357S,2002A&A...387..520P,2007A&A...469..843P}, the
leaking of the edge of disc is realized after several rotation periods
for our simulation.
Although it prompts decrease of the charge from the system
even if all particles are trapped in closed magnetic surface or pair
creation is suppressed, but the speed of the growth of the disc decreases
when the disc grows in vicinity of the light cylinders and the decrease
of charge  is trivial compared with total charge of the disc.
We simulated to confirm how the diocotron instability affects the global
structure of the disc over $20$ rotation periods for the pair starved
electrosphere, which is barely acceptable to carry out for our present
environment.
At least we realized that the diffusion by the diocotron instability was
not so significant in such a time scale, 
although the diffusion by 
$\boldsymbol{f}_{\text{rad}}\times \boldsymbol{B}$ changes the quiet
electrosphere in several rotation periods.
Note that the growth of the disc is made by the pair created positive
charges in the outer gaps and the main component of leaking of the disc
is made by $\boldsymbol{f}_{\text{rad}}\times \boldsymbol{B}$ drift of
the particle in our simulation and even the loss of the positive charge
from the edge of the disc is much lower than the outflow from the outer
gaps to EDR (See, left panel of Fig. \ref{fig:Py9yk6wA},
$a_0\rightarrow a_1$).

If copious plasma is supplied to the equatorial disc from the outer
gaps, another important issue remains the region, the so-called
Y-point, is expected to have the electric field larger than the magnetic
field in the force free theory \citep{2003ApJ...598..446U}.
In the next paper, we treat higher rates of pair creation, for which
modification of the magnetic field becomes significant in the vicinity
of the light cylinder.
Then, some poloidal magnetic flux is opened and toroidal magnetic field
would be anti-parallel on both sides of the equatorial plane in which
the dissipation of the magnetic field is expected to accelerate the
particles by dissipation process of the magnetic field.
Thus, trans-field leakage of the particle around the Y-point
is a very interesting issue, but we postpone further discussions until
the simulations with higher rates of pair creation are performed.

\subsection{Formation process of steady solution}
\label{subsec:8yLnL8je}
Although the steady solution is stated in section \ref{subsec:Ued6jM2g},
here we discuss by the stage of the active solution.
Once pair creation is on set, the pair plasmas are generated at first in
middle latitudes in which $E_{\parallel}$ is stronger than
$E_{\text{cr}}$ (see Fig. \ref{fig:e2caJC6c}).
The pairs created in the outer gaps are immediately separated in
oppsite directions along a magnetic field line by $E_{\parallel}$.
Then, most of the positive charges move into EDR to form an outflow. 
At the same time, the negative charges move inward. 
They returns to the star and are re-emitted from the polar regions so
that the height of the dome increases.
The outflow of the positive particles causes significant change to the
global structure: the system charge, which is initially assumed to 
be positive, is reduced.
As a result, the potential of the polar region becomes negative, and a
part with lower latitude of the polar domes becomes outflow.
The other half of the polar dome grows and crosses over the light
cylinder.
The Lorentz factor of the plasma increases to bring about trans-field
motion by radiation drag force. 
It is notable that the negative charges obtain kinetic energy if they
drift toward inner magnetic surfaces.
Some part of the particles become outflow if the kinetic energy is
larger than the potential energy at the outer boundary, and therefore
the separatrix of the negative charged flow is made on the equator with
$3R_{\text{l}}$ in present result although it is set on the equator
about with $8R_{\text{l}}$ in previous work.
The difference might be made by the resolution of the charge and the
detailed mechanism of the formation of the separatrix with modification
of magnetic field needs additional simulation in future work.
Thus, steady state is found in equal losses of both particle species and
the lost particles are supplied by the outer gaps.

In our simulation, we demonstrate the generation of the particles
intermittently in some periods 
$\bar{\tau}_{\text{pr}}=0.1, 0.2, 1.0, 2.0$ for the pair creation and a
fixed period $\bar{\tau}_{\text{po}}=0.1$ for the popping from the
stellar surface.
All steady results had the same
geometry of the charge clouds, which were equatorial disc of positive
charge, polar domes of negative charge and both signed outflows.
Thus, the choice of $\bar{\tau}_{\text{pr}}$ and
$\bar{\tau}_{\text{po}}$ in present simulation does not affect the
results.
We checked that if the lower system charge solution without pair
creation is chosen as a initial condition, then the same result was
obtained qualitatively.

\subsection{Structure of Outer Gap}
\label{subsec:dcY1gL4y}
We here discuss the structure of the outer gaps for our results.
For the discussion, the structure of the electrostatic potential and the
intensity of $E_{\parallel}$ in the pair creating region should be
realized for our active solutions, where we obtained the solution with
the several rates of the pair creation in simulation.
Note that in our simulation, the $E_{\text{cr}}$ is taken to be large
artificially compared with a parameter for emission of gamma-ray to
reality, in other words, pair creation rate is underestimated.
For the reason, the size of the pair creating region is expanded and
therefore the geometry of the outer gaps is not revealed quantitatively
from our simulation.
Although the structure of the outer gaps which are demonstrated the
lower rate of the pair creation have contained some fraction of
artifact, the result implies how the transition of the
magnetosphere from the active state to the quiet state is realized with
time for the pulsar, that is the evolution of the magnetosphere with
the abundance of the pair plasma, which decreases with age of the
pulsar.

\begin{figure}
 \begin{center}
  \includegraphics[width=8cm]{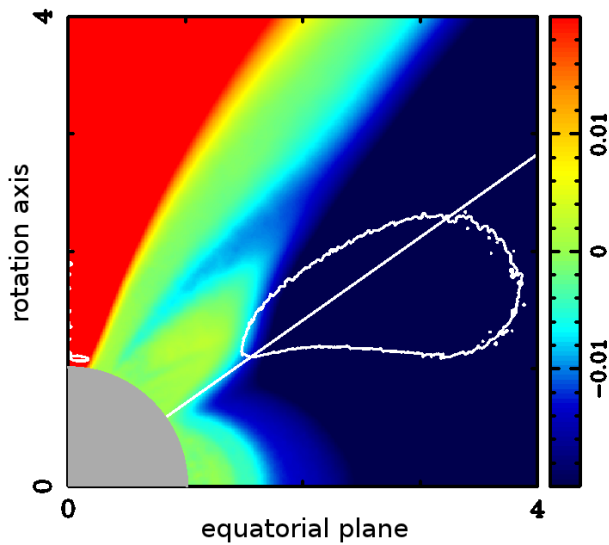}
 \end{center}
 \vspace{-10pt}
 \caption[]{\small{Distribution of non-corotational potential on
 meridional plane, and contour indicates isoline with
 $E_{\parallel}=E_{\text{c}}$. The color of map is normalized by $2\%$
 of the  open field line voltage $\phi_{\text{eff}}$. The line
 in the middle latitude is null surface.}}
 \label{fig:3xbIe5qO}
\end{figure}

Another important point, we can suggest the possibility of
coexistence for polar cap and outer gap 
by comparing with the deviation from the force-free magnetosphere.
If a force-free condition are satisfied elsewhere, we give the
co-rotational potential under dipole magnetic field assumption,
\begin{align} 
 \phi_{\text{co}}=\frac{\Omega \mu \sin^2\theta}{cr}
 \text{.}
 \label{eq:Jx8jD8ls}
\end{align} 
Then, the non-corotational potential defined 
$\phi_{\text{nco}}\equiv \phi-\phi_{\text{co}}$.
Fig. \ref{fig:3xbIe5qO} shows the non-corotational potential with color
map and pair creating region with contour for $B_0$ model on the
meridional plane.
The pair creating region become dramatically small compared with initial
state in our simulation but the intensity of the $E_{\parallel}$ in
steady state remains larger than $E_{\text{cr}}$ in the countour. 
Eventually, the pair creating regions remained in the
middle latitudes and just above the poles.
We don't distinguish between the process of the
pair creations in present simulation,
which are photon-photon collision and magnetic pair creation, and
therefore the rate of the pair creation is only proportional
to the product of the intensity of the $E_{\parallel}$ and the volume of
the gap.
Thus, the supply of the plasma generated in the polar pair creating
region was trivial compared with the one from the outer
gaps, that is, our simulation underestimated the effect of shielding by
pair plasma with magnetic pair creation in the polar cap.
As \cite*{2010ApJ...715.1318T} suggested that a new outer gap closure
mechanism by the magnetic pair creation near the stellar surface is
significant to realize the observed features of the gamma-ray pulsars,
the effect of the different type of the pair creation should be
investigated in future work.

In Fig. \ref{fig:3xbIe5qO}, the equatorial disc and polar domes has
domain in which $\phi_{\text{nco}}$ have deviation less than
$0.01\phi_{\text{eff}}$.
They are not accelerating region of the particle; dead zone.
For our result, the existence of the dead zone provides a inspiration. 
Usually conventional polar cap and outer gap are defined on the
last closed field line for GJ model and the poloidal current in the gaps
directs oppositely, and therefore the gaps can not coexist.
In present result, the dead zone in equatorial disc is reduced in the light
cylinder and the effective last closed line is migrated to higher
colatitude and the one in polar dome cut down the radius of the cone in
which particle outflows.
Note that the dead zones divide the last closed field line and would
be able to coexist polar cap and outer gap on different magnetic
surfaces.
The poloidal current travels into the polar region of the
stellar surface of the lower colatitude side of the polar dead zone,
migrates to the magnetic surface with higher colatitude, and
passes through the outer gaps.
The current makes poloidal circulation beyond the light cylinder.
Because we do not demonstrate detailed processes of pair creation
and do not simulate with enough accuracy to resolve the size of the
polar caps in present results and therefore it is not clear whether
current from the polar cap is necessary condition or not. 
This interesting result should be discussed in future work.

\begin{figure}
 \begin{center}
  \includegraphics[width=8cm]{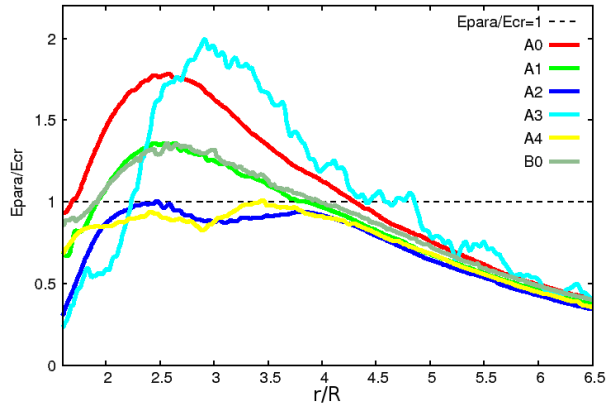}
 \end{center}
 \vspace{-10pt}
 \caption[]{\small{The curves are $E_{\parallel}$ normalized by
 $E_{\text{cr}}$ on the null surface. The horizontal axis indicates
 distance normalized by the stellar radius along the null surface.
 The tags indicate type of simulation referred in Table \ref{table:8yrP4fiZ}.}}
 \label{fig:Kx0pJ0bp}
\end{figure}

For confinement for the robustness of our gap-wind solution, we carried
out the models of some rates of the pair creation.
We checked that $E_{\parallel}$ in the gap saturates with a constant
value for all steady states.
Fig. \ref{fig:Kx0pJ0bp} shows the saturated intensities of $E_{\parallel}$
on the null surface in all simulations.
All results had regions whose $E_{\parallel}$ was larger than
$E_{\text{cr}}$ on the null surface.
The width of the base of the bell-curve on the line with
$E_{\parallel}/E_{\text{cr}}=1$ indicates typical size of the pair
creating region on null surface.
There is a tendency for the higher pair creation rate model
($A_0\rightarrow A_1\rightarrow A_2 \rightarrow A_3$), which is
parameterized by the frequency of the pair creation
$\bar{t}_{\text{pr}}$, to decrease the size of the pair creating region
and the intensity of $E_{\parallel}$.
The saturated intensities are maintained just above the $E_{\text{cr}}$.
When $E_{\text{cr}}$ is taken to be much smaller value 
($A_0\rightarrow A_4 \rightarrow A_5$), the pair creating region
decreased in the same way and the saturated electric field intensity
was held just higher than $E_{\text{cr}}$.
In the lower $E_{\text{cr}}$ model, the particles tended to screen out
the gaps near the star although the outer edge of the pair creating
region does not change.
This indicates that the inner boundary of the outer gap, which a has
beak-like shape on the meridional plane, would not be connected to the
star if we could carry out a much lower $E_{\text{cr}}$ model.
Considering from figure \ref{fig:Py9yk6wA}, the inner edge of the outer
gap would be refined just outside the outer edge of the dead zone which
is roughly defined as a magnetic field line footed on the stellar
surface with $32^{\circ}$ in our simulation.
The inner boundary is slightly inside compared with that of the
conventional outer gap model on the null surface.
This agrees with \cite*{2004MNRAS.354.1120T} which is calculated the
electrodynamics of an outer gap on the meridional plane with assumed
external current, then as the current density increases, the inner
boundary of the outer gap shifts toward the stellar surface.
As previously explained, the main path of equatorial outflow exists
outside of the dead zone, outer gap structure in our simulation can be
maintained steadily.

We emphasize that the thickness of the pair creating region
transversing the dipole magnetic surface would be related to the size of
the region in which leakage of the disc particles takes place 
($\sim f R_l$) because the field aligned potential drop in the outer gap
controls the super-rotation of the disc.
Phenomenologically-postulated, the thickness of the outer gaps for old
pulsars are shown by \cite{1997ApJ...487..370Z} in which they pointed
out the effect of the finite mean free path of the gamma-ray for
pair creation is important for defining the thickness of outer gap.
In addition to this, the direction of the gamma-ray would be important
to define the geometry of the outer gap because the particle has fast
azimuthal motion in the gaps near the light cylinder, and therefore the
gamma-ray is emitted to the azimuthal direction and the path of the
gamma-ray should be considered with three dimensional geometry.
We should consider the mean free path of gamma-ray for pair creation
with a three-dimensional model in future work.

\subsection{Polar Gap}
\label{subsec:c45eshNS}
The polar cap accelerator has been an outstanding issue
in the context of how one can understand radio emissions.
We find a potential drop just above the stellar surface in our
simulation in a similar fashion to the WS.
The mechanism is very simple:
in a steady state the negative charges created in the outer gap flow
back to the star, and therefore the equal amount of negative charges
should be emitted for steadiness.
Then, he potential drop in the polar region of the stellar surface
controls the polar flow of negative particles, which is represented by
\begin{align} 
 \phi (\boldsymbol{R})=
 \frac{\mu\Omega}{cR}
 \left(
 \sin ^2\theta-\frac{2}{3}
 \right)+
 \frac{Q_{\text{sys}}}{R}-
 \sum_{i=1}^n\frac{q_i}{R}\left(1-\frac{R}{r_i}\right)
 \text{.}
 \label{eq:Hrl35soD}
\end{align} 
In our all results of simulations, $\phi_{\text{pole}}$
becomes slightly negative, and therefore the negative particles are
pushed and easily escape to infinity.
Implied by \eqref{eq:Hrl35soD}, this potential is kept by balance
of the numbers of negative and positive particles in the
magnetosphere (the second term) and $Q_{\text{sys}}$ (the third term).
Thus, we see that in the steady state $Q_{\text{sys}}$ is determined so
that the losses of negative and positive particles are balanced. 
This indicates that the polar caps have a close
correlation to the injection of the electron from the the outer gaps.
In the present case, the rate of the emitted particles from the
polar region is less than would be capable of shielding the electric
field above the pole, that is the current density from the pole is less
than $\rho_{\text{GJ}}c$, and therefore the unscreened potential
increases gradually along polar magnetic field lines.
In other words, this potential drop above the pole is not a conventional
polar cap because the electric field is not shielded in finite distance
near the stellar surface.
There is a very interesting issue whether a potential drop near the polar
region remains if enough electrons return from the
outer gaps and are re-emitted from the pole or generated in the polar
cap and then the emitted particles should form the space charge limited
flow \citep{1979ApJ...231..854A}, but our simulation particle has large
inertia length with $3\%$ of stellar radius, and therefore the detailed
structure might not be demonstrated.
Note that our simulation contains the artificial polar gap arising
from the intermittent emission of particle on the stellar surface, which
is in the range of $t_{\text{po}}c=0.1R$ at a maximum.
Unfortunately, it is difficult to take a much smaller value of
$t_{\text{po}}$ in the present parameter setting.
Thus, it is not clear that polar electric field is shielded on the pole if
pair creation process is considered in detail and whether the polar
electric flow needs pair creation in polar cap or not.
As shown by \citep{2003ApJ...591..334H}, the current from polar cap have
a significant effect on outer gap electrodynamics,
to discuss the possibility of existence of the polar caps linked with
the outer gaps, the combination of localized simulation and global simulation
would be needed.

\section{Conclusion}
For an axisymmetric pulsar magnetosphere, pair creation in the outer gaps
results in expansion of rotating electrosphere and
the trans-field drift motion due to radiation drag force near and
beyond the light cylinder. 
Eventually, a steady state is achieved with the outflow of both particle
species and global current loop on the meridional plane.
We confirmed the pair starved static electrosphere shifts to the steady
structure by the pair plasma in the gaps.
The pair creation plays important role of metamorphosing from the static
electrosphere to the steady structure nevertheless the pair
creation rate is artificially suppressed in our present model.
For more detailed discussion of the structure of the magnetosphere,
the consideration of the mean free path of the pair
creation, which is important for old pulsars having thick
outer gaps, is needed to discuss the structure of the gap
quantitatively.
Additionally, a higher pair creation rate case should be simulated, then
modification of the magnetic field from the dipolar changes the pattern
of the outflow of the plasma and the structure of the gaps.
Thus, our model can be developed to consider pulsar magnetosphere
including much more pair plasma in the next paper.

\section*{Acknowledgements}
We would like to thank Jumpei Takata, Hiroyuki Takahashi, Syota
Kisaka and anonymous referee for discussions and many helpful comments
on the manuscript.
Numerical simulations were carried out on the MUV system at Center for
Computational Astrophysics, National Astronomical Observatory
Japan. 


\begin{thebibliography}{51}
\expandafter\ifx\csname natexlab\endcsname\relax\def\natexlab#1{#1}\fi

\bibitem[{{Abdo} {et~al.}(2009{\natexlab{a}}){Abdo}, {Ackermann}, {Ajello},
  {Anderson}, {Atwood}, {Axelsson}, {Baldini}, {Ballet}, {Barbiellini},
  {Baring}, {Bastieri}, {Baughman}, {Bechtol}, {Bellazzini}, {Berenji},
  {Bignami}, {Blandford}, {Bloom}, {Bonamente}, {Borgland}, {Bregeon}, {Brez},
  {Brigida}, {Bruel}, {Burnett}, {Caliandro}, {Cameron}, {Caraveo},
  {Casandjian}, {Cecchi}, {{\c C}elik}, {Chekhtman}, {Cheung}, {Chiang},
  {Ciprini}, {Claus}, {Cohen-Tanugi}, {Conrad}, {Cutini}, {Dermer}, {de
  Angelis}, {de Luca}, {de Palma}, {Digel}, {Dormody}, {do Couto e Silva},
  {Drell}, {Dubois}, {Dumora}, {Farnier}, {Favuzzi}, {Fegan}, {Fukazawa},
  {Funk}, {Fusco}, {Gargano}, {Gasparrini}, {Gehrels}, {Germani}, {Giebels},
  {Giglietto}, {Giommi}, {Giordano}, {Glanzman}, {Godfrey}, {Grenier},
  {Grondin}, {Grove}, {Guillemot}, {Guiriec}, {Gwon}, {Hanabata}, {Harding},
  {Hayashida}, {Hays}, {Hughes}, {J{\'o}hannesson}, {Johnson}, {Johnson},
  {Johnson}, {Kamae}, {Katagiri}, {Kataoka}, {Kawai}, {Kerr}, {Kn{\"o}dlseder},
  {Kocian}, {Kuss}, {Lande}, {Latronico}, {Lemoine-Goumard}, {Longo},
  {Loparco}, {Lott}, {Lovellette}, {Lubrano}, {Madejski}, {Makeev}, {Marelli},
  {Mazziotta}, {McConville}, {McEnery}, {Meurer}, {Michelson}, {Mitthumsiri},
  {Mizuno}, {Monte}, {Monzani}, {Morselli}, {Moskalenko}, {Murgia}, {Nolan},
  {Norris}, {Nuss}, {Ohsugi}, {Omodei}, {Orlando}, {Ormes}, {Paneque},
  {Parent}, {Pelassa}, {Pepe}, {Pesce-Rollins}, {Pierbattista}, {Piron},
  {Porter}, {Primack}, {Rain{\`o}}, {Rando}, {Ray}, {Razzano}, {Rea}, {Reimer},
  {Reimer}, {Reposeur}, {Ritz}, {Rochester}, {Rodriguez}, {Romani}, {Ryde},
  {Sadrozinski}, {Sanchez}, {Sander}, {Parkinson}, {Scargle}, {Sgr{\`o}},
  {Siskind}, {Smith}, {Smith}, {Spandre}, {Spinelli}, {Starck}, {Strickman},
  {Suson}, {Tajima}, {Takahashi}, {Takahashi}, {Tanaka}, {Thayer}, {Thompson},
  {Tibaldo}, {Tibolla}, {Torres}, {Tosti}, {Tramacere}, {Uchiyama}, {Usher},
  {Van Etten}, {Vasileiou}, {Vilchez}, {Vitale}, {Waite}, {Wang}, {Watters},
  {Winer}, {Wolff}, {Wood}, {Ylinen}, \& {Ziegler}}]{2009Sci...325..840A}
{Abdo} A.~A. et~al., Science, 325, 840

\bibitem[{{Abdo} {et~al.}(2009{\natexlab{b}}){Abdo}, {Ackermann}, {Ajello},
  {Atwood}, {Axelsson}, {Baldini}, {Ballet}, {Barbiellini}, {Baring},
  {Bastieri}, {Baughman}, {Bechtol}, {Bellazzini}, {Berenji}, {Bignami},
  {Blandford}, {Bloom}, {Bonamente}, {Borgland}, {Bregeon}, {Brez}, {Brigida},
  {Bruel}, {Burnett}, {Caliandro}, {Cameron}, {Camilo}, {Caraveo}, {Carlson},
  {Casandjian}, {Cecchi}, {{\c C}elik}, {Charles}, {Chekhtman}, {Cheung},
  {Chiang}, {Ciprini}, {Claus}, {Cognard}, {Cohen-Tanugi}, {Cominsky},
  {Conrad}, {Corbet}, {Cutini}, {Dermer}, {Desvignes}, {de Angelis}, {de Luca},
  {de Palma}, {Digel}, {Dormody}, {do Couto e Silva}, {Drell}, {Dubois},
  {Dumora}, {Edmonds}, {Farnier}, {Favuzzi}, {Fegan}, {Focke}, {Frailis},
  {Freire}, {Fukazawa}, {Funk}, {Fusco}, {Gargano}, {Gasparrini}, {Gehrels},
  {Germani}, {Giebels}, {Giglietto}, {Giordano}, {Glanzman}, {Godfrey},
  {Grenier}, {Grondin}, {Grove}, {Guillemot}, {Guiriec}, {Hanabata}, {Harding},
  {Hayashida}, {Hays}, {Hobbs}, {Hughes}, {J{\'o}hannesson}, {Johnson},
  {Johnson}, {Johnson}, {Johnson}, {Johnston}, {Kamae}, {Katagiri}, {Kataoka},
  {Kawai}, {Kerr}, {Kn{\"o}dlseder}, {Kocian}, {Kramer}, {Kuss}, {Lande},
  {Latronico}, {Lemoine-Goumard}, {Longo}, {Loparco}, {Lott}, {Lovellette},
  {Lubrano}, {Madejski}, {Makeev}, {Manchester}, {Marelli}, {Mazziotta},
  {McConville}, {McEnery}, {McLaughlin}, {Meurer}, {Michelson}, {Mitthumsiri},
  {Mizuno}, {Moiseev}, {Monte}, {Monzani}, {Morselli}, {Moskalenko}, {Murgia},
  {Nolan}, {Norris}, {Nuss}, {Ohsugi}, {Omodei}, {Orlando}, {Ormes}, {Paneque},
  {Panetta}, {Parent}, {Pelassa}, {Pepe}, {Pesce-Rollins}, {Piron}, {Porter},
  {Rain{\`o}}, {Rando}, {Ransom}, {Ray}, {Razzano}, {Rea}, {Reimer}, {Reimer},
  {Reposeur}, {Ritz}, {Rochester}, {Rodriguez}, {Romani}, {Roth}, {Ryde},
  {Sadrozinski}, {Sanchez}, {Sander}, {Saz Parkinson}, {Scargle}, {Schalk},
  {Sgr{\`o}}, {Siskind}, {Smith}, {Smith}, {Spandre}, {Spinelli}, {Stappers},
  {Starck}, {Striani}, {Strickman}, {Suson}, {Tajima}, {Takahashi}, {Tanaka},
  {Thayer}, {Thayer}, {Theureau}, {Thompson}, {Thorsett}, {Tibaldo}, {Torres},
  {Tosti}, {Tramacere}, {Uchiyama}, {Usher}, {Van Etten}, {Vasileiou},
  {Venter}, {Vilchez}, {Vitale}, {Waite}, {Wallace}, {Wang}, {Watters}, {Webb},
  {Weltevrede}, {Winer}, {Wood}, {Ylinen}, \& {Ziegler}}]{2009Sci...325..848A}
{Abdo} A.~A. et~al., 2009{\natexlab{b}}, Science, 325, 848

\bibitem[{{Abdo} {et~al.}(2010){Abdo}, {Ackermann}, {Ajello}, {Atwood},
  {Axelsson}, {Baldini}, {Ballet}, {Barbiellini}, {Baring}, {Bastieri},
  {Baughman}, {Bechtol}, {Bellazzini}, {Berenji}, {Blandford}, {Bloom},
  {Bonamente}, {Borgland}, {Bregeon}, {Brez}, {Brigida}, {Bruel}, {Burnett},
  {Buson}, {Caliandro}, {Cameron}, {Camilo}, {Caraveo}, {Casandjian}, {Cecchi},
  {{\c C}elik}, {Charles}, {Chekhtman}, {Cheung}, {Chiang}, {Ciprini}, {Claus},
  {Cognard}, {Cohen-Tanugi}, {Cominsky}, {Conrad}, {Corbet}, {Cutini}, {den
  Hartog}, {Dermer}, {de Angelis}, {de Luca}, {de Palma}, {Digel}, {Dormody},
  {Silva}, {Drell}, {Dubois}, {Dumora}, {Espinoza}, {Farnier}, {Favuzzi},
  {Fegan}, {Ferrara}, {Focke}, {Fortin}, {Frailis}, {Freire}, {Fukazawa},
  {Funk}, {Fusco}, {Gargano}, {Gasparrini}, {Gehrels}, {Germani}, {Giavitto},
  {Giebels}, {Giglietto}, {Giommi}, {Giordano}, {Glanzman}, {Godfrey},
  {Gotthelf}, {Grenier}, {Grondin}, {Grove}, {Guillemot}, {Guiriec}, {Gwon},
  {Hanabata}, {Harding}, {Hayashida}, {Hays}, {Hughes}, {Jackson},
  {J{\'o}hannesson}, {Johnson}, {Johnson}, {Johnson}, {Johnson}, {Johnston},
  {Kamae}, {Kanbach}, {Kaspi}, {Katagiri}, {Kataoka}, {Kawai}, {Kerr},
  {Kn{\"o}dlseder}, {Kocian}, {Kramer}, {Kuss}, {Lande}, {Latronico},
  {Lemoine-Goumard}, {Livingstone}, {Longo}, {Loparco}, {Lott}, {Lovellette},
  {Lubrano}, {Lyne}, {Madejski}, {Makeev}, {Manchester}, {Marelli},
  {Mazziotta}, {McConville}, {McEnery}, {McGlynn}, {Meurer}, {Michelson},
  {Mineo}, {Mitthumsiri}, {Mizuno}, {Moiseev}, {Monte}, {Monzani}, {Morselli},
  {Moskalenko}, {Murgia}, {Nakamori}, {Nolan}, {Norris}, {Noutsos}, {Nuss},
  {Ohsugi}, {Omodei}, {Orlando}, {Ormes}, {Ozaki}, {Paneque}, {Panetta},
  {Parent}, {Pelassa}, {Pepe}, {Pesce-Rollins}, {Piron}, {Porter}, {Rain{\`o}},
  {Rando}, {Ransom}, {Ray}, {Razzano}, {Rea}, {Reimer}, {Reimer}, {Reposeur},
  {Ritz}, {Rodriguez}, {Romani}, {Roth}, {Ryde}, {Sadrozinski}, {Sanchez},
  {Sander}, {Saz Parkinson}, {Scargle}, {Schalk}, {Sellerholm}, {Sgr{\`o}},
  {Siskind}, {Smith}, {Smith}, {Spandre}, {Spinelli}, {Stappers}, {Starck},
  {Striani}, {Strickman}, {Strong}, {Suson}, {Tajima}, {Takahashi},
  {Takahashi}, {Tanaka}, {Thayer}, {Thayer}, {Theureau}, {Thompson},
  {Thorsett}, {Tibaldo}, {Tibolla}, {Torres}, {Tosti}, {Tramacere}, {Uchiyama},
  {Usher}, {Van Etten}, {Vasileiou}, {Venter}, {Vilchez}, {Vitale}, {Waite},
  {Wang}, {Wang}, {Watters}, {Weltevrede}, {Winer}, {Wood}, {Ylinen}, \&
  {Ziegler}}]{2010ApJS..187..460A}
{Abdo} A.~A. et al., 2010, \apjs, 187, 460


\bibitem[{{Abdo} {et~al.}(2009{\natexlab{c}}){Abdo}, {Ackermann}, {Atwood},
  {Bagagli}, {Baldini}, {Ballet}, {Band}, {Barbiellini}, {Baring}, {Bartelt},
  {Bastieri}, {Baughman}, {Bechtol}, {Bellardi}, {Bellazzini}, {Berenji},
  {Bisello}, {Blandford}, {Bloom}, {Bogart}, {Bonamente}, {Borgland},
  {Bouvier}, {Bregeon}, {Brez}, {Brigida}, {Bruel}, {Burnett}, {Caliandro},
  {Cameron}, {Camilo}, {Caraveo}, {Casandjian}, {Ceccanti}, {Cecchi},
  {Charles}, {Chekhtman}, {Cheung}, {Chiang}, {Ciprini}, {Claus}, {Cognard},
  {Cohen-Tanugi}, {Cominsky}, {Conrad}, {Corbet}, {Corucci}, {Cutini}, {Davis},
  {DeKlotz}, {Dermer}, {de Angelis}, {de Palma}, {Digel}, {Dormody}, {Silva},
  {Drell}, {Dubois}, {Dumora}, {Espinoza}, {Farnier}, {Favuzzi}, {Flath},
  {Fleury}, {Focke}, {Frailis}, {Friere}, {Fukazawa}, {Funk}, {Fusco},
  {Gargano}, {Gasparrini}, {Gehrels}, {Germani}, {Giannitrapani}, {Giebels},
  {Giglietto}, {Giordano}, {Glanzman}, {Godfrey}, {Gotthelf}, {Grenier},
  {Grondin}, {Grove}, {Guillemot}, {Guiriec}, {Haller}, {Harding}, {Hart},
  {Hartman}, {Hays}, {Hobbs}, {Hughes}, {J{\'o}hannesson}, {Johnson},
  {Johnson}, {Johnson}, {Johnson}, {Johnston}, {Kamae}, {Kanbach}, {Kaspi},
  {Katagiri}, {Kataoka}, {Kavelaars}, {Kawai}, {Kelly}, {Kerr}, {Kiziltan},
  {Klamra}, {Kn{\"o}dlseder}, {Kramer}, {Kuehn}, {Kuss}, {Lande}, {Landriu},
  {Latronico}, {Lee}, {Lee}, {Lemoine-Goumard}, {Livingstone}, {Longo},
  {Loparco}, {Lott}, {Lovellette}, {Lubrano}, {Lyne}, {Madejski}, {Makeev},
  {Manchester}, {Marangelli}, {Marelli}, {Mazziotta}, {McEnery}, {McGlynn},
  {McLaughlin}, {Menon}, {Meurer}, {Michelson}, {Mineo}, {Mirizzi},
  {Mitthumsiri}, {Mizuno}, {Moiseev}, {Mongelli}, {Monte}, {Monzani},
  {Moretti}, {Morselli}, {Moskalenko}, {Murgia}, {Nakamori}, {Nolan},
  {Noutsos}, {Nuss}, {Ohsugi}, {Omodei}, {Orlando}, {Ormes}, {Ozaki},
  {Paccagnella}, {Paneque}, {Panetta}, {Parent}, {Pearce}, {Pepe},
  {Perchiazzi}, {Pesce-Rollins}, {Pieri}, {Pinchera}, {Piron}, {Porter},
  {Rain{\`o}}, {Rando}, {Ransom}, {Rapposelli}, {Razzano}, {Reimer}, {Reimer},
  {Reposeur}, {Reyes}, {Ritz}, {Rochester}, {Rodriguez}, {Romani}, {Roth},
  {Ryde}, {Sacchetti}, {Sadrozinski}, {Saggini}, {Sanchez}, {Sander},
  {Parkinson}, {Segal}, {Sellerholm}, {Sgr{\`o}}, {Siskind}, {Smith}, {Smith},
  {Spandre}, {Spinelli}, {Stamatikos}, {Starck}, {Stecker}, {Stephens},
  {Strickman}, {Strong}, {Suson}, {Tajima}, {Takahashi}, {Takahashi}, {Tanaka},
  {Tenze}, {Thayer}, {Thayer}, {Theureau}, {Thompson}, {Thorsett}, {Tibaldo},
  {Tibolla}, {Torres}, {Tramacere}, {Turri}, {Usher}, {Vigiani}, {Vilchez},
  {Vitale}, {Waite}, {Wang}, {Watters}, {Weltevrede}, {Winer}, {Wood},
  {Ylinen}, \& {Ziegler}}]{2009ApJ...696.1084A}
{Abdo} A.~A. et~al., 2009{\natexlab{c}}, \apj, 696, 1084

\bibitem[{{Aliu} {et~al.}(2008){Aliu}, {Anderhub}, {Antonelli}, {Antoranz},
  {Backes}, {Baixeras}, {Barrio}, {Bartko}, {Bastieri}, {Becker}, {Bednarek},
  {Berger}, {Bernardini}, {Bigongiari}, {Biland}, {Bock}, {Bonnoli}, {Bordas},
  {Bosch-Ramon}, {Bretz}, {Britvitch}, {Camara}, {Carmona}, {Chilingarian},
  {Commichau}, {Contreras}, {Cortina}, {Costado}, {Covino}, {Curtef}, {Dazzi},
  {De Angelis}, {De Cea del Pozo}, {de los Reyes}, {De Lotto}, {De Maria}, {De
  Sabata}, {Delgado Mendez}, {Dominguez}, {Dorner}, {Doro}, {Els{\"a}sser},
  {Errando}, {Fagiolini}, {Ferenc}, {Fernandez}, {Firpo}, {Fonseca}, {Font},
  {Galante}, {Garcia Lopez}, {Garczarczyk}, {Gaug}, {Goebel}, {Hadasch},
  {Hayashida}, {Herrero}, {H{\"o}hne}, {Hose}, {Hsu}, {Huber}, {Jogler},
  {Kranich}, {La Barbera}, {Laille}, {Leonardo}, {Lindfors}, {Lombardi},
  {Longo}, {Lopez}, {Lorenz}, {Majumdar}, {Maneva}, {Mankuzhiyil}, {Mannheim},
  {Maraschi}, {Mariotti}, {Martinez}, {Mazin}, {Meucci}, {Meyer}, {Miranda},
  {Mirzoyan}, {Moles}, {Moralejo}, {Nieto}, {Nilsson}, {Ninkovic}, {Otte},
  {Oya}, {Paoletti}, {Paredes}, {Pasanen}, {Pascoli}, {Pauss}, {Pegna},
  {Perez-Torres}, {Persic}, {Peruzzo}, {Piccioli}, {Prada}, {Prandini},
  {Puchades}, {Raymers}, {Rhode}, {Rib{\'o}}, {Rico}, {Rissi}, {Robert},
  {R{\"u}gamer}, {Saggion}, {Saito}, {Salvati}, {Sanchez-Conde}, {Sartori},
  {Satalecka}, {Scalzotto}, {Scapin}, {Schweizer}, {Shayduk}, {Shinozaki},
  {Shore}, {Sidro}, {Sierpowska-Bartosik}, {Sillanp{\"a}{\"a}}, {Sobczynska},
  {Spanier}, {Stamerra}, {Stark}, {Takalo}, {Tavecchio}, {Temnikov}, {Tescaro},
  {Teshima}, {Tluczykont}, {Torres}, {Turini}, {Vankov}, {Venturini}, {Vitale},
  {Wagner}, {Wittek}, {Zabalza}, {Zandanel}, {Zanin}, {Zapatero}, {de Jager},
  {de Ona Wilhelmi}, \& {MAGIC Collaboration}}]{2008Sci...322.1221A}
{Aliu} E. et al,  {MAGIC Collaboration}, 2008, Science, 322, 1221

\bibitem[{{Arons} \& {Scharlemann}(1979)}]{1979ApJ...231..854A}
{Arons} J., {Scharlemann} E.~T., 1979, \apj, 231, 854

\bibitem[{{Becker} \& {Tr\"{u}emper}(1997)}]{1997A&A...326..682B}
{Becker} W., {Tr\"{u}emper} J., 1997, \aap, 326, 682

\bibitem[{{Bucciantini} {et~al.}(2006){Bucciantini}, {Thompson}, {Arons},
  {Quataert}, \& {Del Zanna}}]{2006MNRAS.368.1717B}
{Bucciantini} N. et al., 2006, \mnras, 368, 1717

\bibitem[{{Cheng} {et~al.}(1986{\natexlab{a}}){Cheng}, {Ho}, \&
  {Ruderman}}]{1986ApJ...300..500C}
{Cheng} K.~S., {Ho} C., {Ruderman} M., 1986{\natexlab{a}}, \apj, 300, 500

\bibitem[{{Cheng} {et~al.}(1986{\natexlab{b}}){Cheng}, {Ho}, \&
  {Ruderman}}]{1986ApJ...300..522C}
---, 1986{\natexlab{b}}, \apj, 300, 522

\bibitem[{{Daugherty} \& {Harding}(1996)}]{1996ApJ...458..278D}
{Daugherty} J.~K., {Harding} A.~K., 1996, \apj, 458, 278

\bibitem[{{de Jager} \& {Djannati-Ata{\"i}}(2008)}]{2008arXiv0803.0116D}
{de Jager} O.~C., {Djannati-Ata{\"i}} A., 2008, ArXiv e-prints, 803

\bibitem[{{Goldreich} \& {Julian}(1969)}]{1969ApJ...157..869G}
{Goldreich} P., {Julian} W.~H., 1969, \apj, 157, 869

\bibitem[{{Harding} \& {Muslimov}(2002)}]{2002ApJ...568..862H}
{Harding} A.~K., {Muslimov} A.~G., 2002, \apj, 568, 862

\bibitem[{{Hirotani} {et~al.}(2003){Hirotani}, {Harding}, \&
  {Shibata}}]{2003ApJ...591..334H}
{Hirotani} K., {Harding} A.~K., {Shibata} S., 2003, \apj, 591, 334

\bibitem[{{Hirotani} \& {Shibata}(1999)}]{1999MNRAS.308...54H}
{Hirotani} K., {Shibata} S., 1999, \mnras, 308, 54

\bibitem[{{Hobbs} {et~al.}(2004){Hobbs}, {Faulkner}, {Stairs}, {Camilo},
  {Manchester}, {Lyne}, {Kramer}, {D'Amico}, {Kaspi}, {Possenti}, {McLaughlin},
  {Lorimer}, {Burgay}, {Joshi}, \& {Crawford}}]{2004MNRAS.352.1439H}
{Hobbs} G. et al., 2004, \mnras, 352, 1439

\bibitem[{{Holloway}(1973)}]{1973Natur.246....6H}
{Holloway} N.~J., 1973, \nat, 246, 6

\bibitem[{{Jackson}(1976)}]{1976ApJ...206..831J}
{Jackson} E.~A., 1976, \apj, 206, 831

\bibitem[{{Kargaltsev} \& {Pavlov}(2008)}]{2008AIPC..983..171K}
{Kargaltsev} O., {Pavlov} G.~G., 2008, in American Institute of Physics
  Conference Series, Vol. 983, American Institute of Physics Conference Series,
  pp. 171--185

\bibitem[{{Kennel} \& {Coroniti}(1984{\natexlab{a}})}]{1984ApJ...283..694K}
{Kennel} C.~F., {Coroniti} F.~V., 1984{\natexlab{a}}, \apj, 283, 694

\bibitem[{{Kennel} \& {Coroniti}(1984{\natexlab{b}})}]{1984ApJ...283..710K}
---, 1984{\natexlab{b}}, \apj, 283, 710

\bibitem[{{Kirk} {et~al.}(2002){Kirk}, {Skj{\ae}raasen}, \&
  {Gallant}}]{2002A&A...388L..29K}
{Kirk} J.~G., {Skj{\ae}raasen} O., {Gallant} Y.~A., 2002, \aap, 388, L29

\bibitem[{{Komissarov}(2006)}]{2006MNRAS.367...19K}
{Komissarov} S.~S., 2006, \mnras, 367, 19

\bibitem[{{Krause-Polstorff} \&
  {Michel}(1985{\natexlab{a}})}]{1985MNRAS.213P..43K}
{Krause-Polstorff} J., {Michel} F.~C., 1985{\natexlab{a}}, \mnras, 213, 43P

\bibitem[{{Krause-Polstorff} \&
  {Michel}(1985{\natexlab{b}})}]{1985A&A...144...72K}
---, 1985{\natexlab{b}}, \aap, 144, 72

\bibitem[{Makino {et~al.}(2007)Makino, Hiraki, \& Inaba}]{Makino2007}
Makino J., Hiraki K., Inaba M., 2007, in Proceedings of the 2007 ACM/IEEE
  conference on Supercomputing-Volume 00, ACM, pp. 1--11

\bibitem[{{McKinney}(2006{\natexlab{a}})}]{2006MNRAS.367.1797M}
{McKinney} J.~C., 2006{\natexlab{a}}, \mnras, 367, 1797

\bibitem[{{McKinney}(2006{\natexlab{b}})}]{2006MNRAS.368L..30M}
---, 2006{\natexlab{b}}, \mnras, 368, L30

\bibitem[{{Michel}(2004)}]{2004AdSpR..33..542M}
{Michel} F.~C., 2004, Advances in Space Research, 33, 542

\bibitem[{{Michel} \& {Smith}(2001)}]{2001RMxAC..10..168M}
{Michel} F.~C., {Smith} I.~A., 2001, in Revista Mexicana de Astronomia y
  Astrofisica Conference Series, Vol.~10, Revista Mexicana de Astronomia y
  Astrofisica Conference Series, {J.~Cant{\'o} \& L.~F.~Rodr{\'{\i}}guez}, ed.,
  pp. 168--175

\bibitem[{{Muslimov} \& {Harding}(2003)}]{2003ApJ...588..430M}
{Muslimov} A.~G., {Harding} A.~K., 2003, \apj, 588, 430

\bibitem[{{Muslimov} \& {Harding}(2004)}]{2004ApJ...606.1143M}
---, 2004, \apj, 606, 1143

\bibitem[{{Nel} {et~al.}(1996){Nel}, {Arzoumanian}, {Bailes}, {Brazier},
  {D'Amico}, {Esposito}, {Fichtel}, {Fierro}, {Hunter}, {Johnston}, {Kanbach},
  {Kaspi}, {Kniffen}, {Lin}, {Lyne}, {Manchester}, {Mattox},
  {Mayer-Hasselwander}, {Merck}, {Michelson}, {Nice}, {Nolan}, {Ramanamurthy},
  {Taylor}, {Thompson}, \& {Westbrook}}]{1996ApJ...465..898N}
{Nel} H.~I. et al., 1996, \apj, 465, 898

\bibitem[{{P{\'e}tri}(2007)}]{2007A&A...469..843P}
{P{\'e}tri} J., 2007, \aap, 469, 843

\bibitem[{{P{\'e}tri} {et~al.}(2002{\natexlab{a}}){P{\'e}tri}, {Heyvaerts}, \&
  {Bonazzola}}]{2002A&A...387..520P}
{P{\'e}tri} J., {Heyvaerts} J., {Bonazzola} S., 2002{\natexlab{a}}, \aap, 387,
  520

\bibitem[{{P{\'e}tri} {et~al.}(2002{\natexlab{b}}){P{\'e}tri}, {Heyvaerts}, \&
  {Bonazzola}}]{2002A&A...384..414P}
---, 2002{\natexlab{b}}, \aap, 384, 414

\bibitem[{{Possenti} {et~al.}(2002){Possenti}, {Cerutti}, {Colpi}, \&
  {Mereghetti}}]{2002A&A...387..993P}
{Possenti} A., {Cerutti} R., {Colpi} M., {Mereghetti} S., 2002, \aap, 387, 993

\bibitem[{{Rees} \& {Gunn}(1974)}]{1974MNRAS.167....1R}
{Rees} M.~J., {Gunn} J.~E., 1974, \mnras, 167, 1

\bibitem[{{Romani}(1996)}]{1996ApJ...470..469R}
{Romani} R.~W., 1996, \apj, 470, 469

\bibitem[{{Ruderman} \& {Sutherland}(1975)}]{1975ApJ...196...51R}
{Ruderman} M.~A., {Sutherland} P.~G., 1975, \apj, 196, 51

\bibitem[{{Rylov}(1977)}]{1977Ap&SS..51...59R}
{Rylov} I.~A., 1977, \apss, 51, 59

\bibitem[{{Smith} {et~al.}(2001){Smith}, {Michel}, \&
  {Thacker}}]{2001MNRAS.322..209S}
{Smith} I.~A., {Michel} F.~C., {Thacker} P.~D., 2001, \mnras, 322, 209

\bibitem[{{Spitkovsky}(2004)}]{2004IAUS..218..357S}
{Spitkovsky} A., 2004, in IAU Symposium, Vol. 218, Young Neutron Stars and
  Their Environments, {F.~Camilo \& B.~M.~Gaensler}, ed., pp. 357--+

\bibitem[{{Takata} {et~al.}(2004){Takata}, {Shibata}, \&
  {Hirotani}}]{2004MNRAS.354.1120T}
{Takata} J., {Shibata} S., {Hirotani} K., 2004, \mnras, 354, 1120

\bibitem[{{Takata} {et~al.}(2006){Takata}, {Shibata}, {Hirotani}, \&
  {Chang}}]{2006MNRAS.366.1310T}
{Takata} J., {Shibata} S., {Hirotani} K., {Chang} H.-K., 2006, \mnras, 366,
  1310

\bibitem[{{Takata} {et~al.}(2010){Takata}, {Wang}, \&
  {Cheng}}]{2010ApJ...715.1318T}
{Takata} J., {Wang} Y., {Cheng} K.~S., 2010, \apj, 715, 1318

\bibitem[{{Uzdensky}(2003)}]{2003ApJ...598..446U}
{Uzdensky} D.~A., 2003, \apj, 598, 446

\bibitem[{{Wada} \& {Shibata}(2007)}]{2007MNRAS.376.1460W}
{Wada} T., {Shibata} S., 2007, \mnras, 376, 1460

\bibitem[{{Zhang} \& {Harding}(2000)}]{2000ApJ...532.1150Z}
{Zhang} B., {Harding} A.~K., 2000, \apj, 532, 1150

\bibitem[{{Zhang} \& {Cheng}(1997)}]{1997ApJ...487..370Z}
{Zhang} L., {Cheng} K.~S., 1997, \apj, 487, 370

\end{thebibliography}

\label{lastpage}
\end{document}